\documentclass[pdftex,twocolumn,epjc3]{svjour3}          

\RequirePackage[T1]{fontenc}

\smartqed  

\RequirePackage{graphicx}
\RequirePackage{mathptmx,amsmath,amssymb,gensymb}      
\RequirePackage{flushend}
\RequirePackage[numbers,sort&compress]{natbib}
\RequirePackage[colorlinks,citecolor=blue,urlcolor=blue,linkcolor=blue]{hyperref}
\RequirePackage{bbding}
\journalname{Eur. Phys. J. C}

\begin{document}

\title{Verifiable type-III seesaw and dark matter  in a gauged ${U(1)_{\rm B-L}}$ symmetric model}

\author{Satyabrata Mahapatra\thanksref{e1,addr1,addr2}
	\and
	Partha Kumar Paul\thanksref{e2,addr3}
	\and
	Narendra Sahu\thanksref{e3,addr3}
	\and
	Prashant Shukla\thanksref{e4,addr4,addr5}
}

\thankstext{e1}{e-mail: satyabrata@iitgoa.ac.in}
\thankstext{e2}{e-mail: ph22resch11012@iith.ac.in}
\thankstext{e3}{e-mail: nsahu@phy.iith.ac.in}
\thankstext{e4}{e-mail: pshukla@barc.gov.in}

\institute{School of Physical Sciences, Indian Institute of Technology Goa, Ponda-403401, Goa, India\label{addr1}
	\and
	Department of Physics and Institute of Basic Science, Sungkyunkwan University, Suwon 16419, Korea\label{addr2}
	\and
	Department of Physics, Indian Institute of Technology Hyderabad, Kandi, Telangana-502285, India\label{addr3}
	\and
	Nuclear Physics Division, Bhabha Atomic Research Centre,
	Mumbai, 400085, India.\label{addr4}
	\and
	Homi Bhabha National Institute, Anushakti Nagar, Mumbai,
	400094, India.\label{addr5}
}

\date{}

\maketitle

\begin{abstract}
We propose a new extension of the Standard Model that incorporates a gauged \( U(1)_{\rm B-L} \) symmetry and the type-III seesaw mechanism to explain neutrino mass generation and provide a viable dark matter (DM) candidate. Unlike the type-I seesaw, the type-III seesaw extension under \( U(1)_{\rm B-L} \) is not automatically anomaly-free. We show that these anomalies can be canceled by introducing additional chiral fermions, which naturally emerge as DM candidates in the model. We thoroughly analyze the DM phenomenology, including relic density, direct and indirect detection prospects, and constraints from current experimental data. Furthermore, we explore the collider signatures of the model, highlighting the enhanced production cross-section of the triplet fermions mediated by the \( \rm B-L \) gauge boson, as well as the potential disappearing track signatures. Additionally, we investigate the gravitational wave signals arising from the first-order phase transition during \( \rm B-L \) symmetry breaking, offering a complementary cosmological probe of the framework.
\end{abstract}

\section{Introduction}
\label{intro}
The Standard Model (SM) of particle physics, despite its remarkable success, leaves several fundamental questions unanswered. Two of the most pressing mysteries are the nature of dark matter (DM) and the origin of neutrino masses. Observational evidence from astrophysics and cosmology, including galaxy rotation curves, gravitational lensing, and cosmic microwave background data, strongly suggests that DM constitutes approximately $85\%$ of the Universe's matter content and $26.8\%$ of its energy density~\cite{Planck:2018vyg,ParticleDataGroup:2024cfk}. However, the SM lacks a suitable candidate to explain the DM of the universe.

Concurrently, the discovery of neutrino oscillations~\cite{Super-Kamiokande:1998kpq,SNO:2001kpb,DoubleChooz:2011ymz,DayaBay:2012fng,RENO:2012mkc} has conclusively demonstrated that neutrinos possess non-zero masses, contradicting the SM prediction of massless neutrinos. This observation necessitates an extension of the SM to accommodate neutrino mass generation mechanisms. Various seesaw mechanisms have been proposed to explain the small neutrino masses, including the type I~\cite{Minkowski:1977sc,Gell-Mann:1979vob,Mohapatra:1979ia,Schechter:1980gr,yanagida}, type II~\cite{Mohapatra:1980yp,Lazarides:1980nt,Wetterich:1981bx,Schechter:1981cv,Ma:1998dx}, and type III~\cite{Foot:1988aq} seesaw models.

The potential connection between these two fundamental puzzles has sparked considerable interest in the particle physics community. Several beyond Standard Model (BSM) theories have been developed to address both issues within a unified framework. Among these, models based on the gauged $U(1)_{\rm B-L}$ symmetry have garnered significant attention due to their simplicity and potential for experimental verifiability. See for instance ~\cite{Davidson:1978pm, Mohapatra:1980qe,Buchmuller:1991ce,Sahu:2005fe,Montero:2007cd,Ma:2014qra,Sanchez-Vega:2015qva,Biswas:2019ygr,Das:2020uer,Mahapatra:2020dgk,Ghosh:2021khk} and the references there in.

In this paper, we present a comprehensive study of a $U(1)_{\rm B-L}$ gauged type-III seesaw model that simultaneously addresses the neutrino mass puzzle and provides a DM candidate. We explore the model's theoretical foundations, analyze its phenomenological consequences, and discuss its implications for current and future experimental searches. The type-III seesaw mechanism~\cite{Foot:1988aq}, which introduces $SU(2)_L$ fermion triplets, offers a unique approach to neutrino mass generation. However, when embedded within a gauged $U(1)_{\rm B-L}$ symmetric setup, it introduces new anomalies that require additional particle content for cancellation. Intriguingly, the fermions needed for anomaly cancellation in this setup can serve as viable DM candidates, establishing a natural link between neutrino physics and DM. Specifically, after a careful assignment of $\rm B-L$ charges to the triplet fermions which successfully cancels the $SU(2)_L^2 \times U(1)_{\rm B-L}$ anomalies, two additional chiral fermions are required to cancel the remaining $U(1)_{\rm B-L}^3$ and $(grav.)^2\times U(1)_{\rm B-L}$ anomalies. These fermions can combine to form a Dirac particle, which gets stabilized by the remnant $Z_2$ symmetry after the $U(1)_{\rm B-L}$ breaking and thus becomes a natural DM candidate.

Here, it is worth noting that, as the triplet fermions carry electroweak charge, their production and decay signatures can be prominent at colliders as compared to the type-I seesaw scenario, and by the virtue of their  $SU(2)_L$ interactions, they can be pair‑produced via electroweak processes with cross sections set by the triplet mass~\cite{Cai:2017mow, Jana:2020qzn}. In contrast, heavy right-handed neutrinos in type-I seesaw are SM singlets and are only produced via tiny mixing angles, making them essentially invisible at colliders. However, a $\Sigma$ can be copiously produced at the LHC/HL-LHC once kinematically allowed. 
The charged component $\Sigma^{\pm}$ decays to the neutral $\Sigma^{0}$ plus soft pions (or leptons) with a potentially long lifetime because of the phase space suppression due to the small mass difference arising because of electroweak corrections~\cite{Cirelli:2005uq}. If $\Sigma^{\pm}$ is moderately long-lived, it yields a disappearing track signature in the detector. This striking signal, absent in type-I models, can be searched for at the LHC/HL-LHC. Type-III models also readily produce displaced vertex events, and proposed detectors like MATHUSLA or future lepton–hadron colliders (LHeC, FCC-he) could detect such displaced decays~\cite{Cai:2017mow, Jana:2020qzn}. Thus, type-III seesaw implementations can be far more testable and constrained than the minimal type-I seesaw.

Furthermore the $U(1)_{\rm B-L}$ gauged type-III seesaw model we present here presents rich phenomenological implications for DM studies. It offers novel avenues for investigating DM relic density, direct and indirect detection strategies, and collider searches. The presence of the $\rm B-L$ breaking scalar helps in obtaining an enhanced parameter space for the DM. Notably, the model predicts unique collider signatures arising from the production of triplet fermions mediated by the $ \rm B-L $ gauge boson. For TeV-scale triplet fermions, the production cross-section at colliders can be significantly enhanced in the presence of a 
$ \rm B-L $ gauge boson with a mass in the few-TeV range. 

In addition, the model also predicts unique cosmological signatures in the form of stochastic gravitational waves (GWs). These GWs originate from first-order phase transitions during the breaking of the $U(1)_{\rm B-L}$ symmetry, a process central to the framework's phenomenology. The resulting GW spectrum can be probed by current and future observatories such as LISA~\cite{LISA:2017pwj}, DECIGO~\cite{Adelberger:2005bt}, and the Einstein Telescope~\cite{Punturo:2010zz}. These experiments provide a complementary avenue to test the model, as the GW signals are sensitive to the dynamics of the phase transition and the underlying particle physics parameters.

In contrast to generic seesaw+DM extensions, a decisive feature of our gauged $U(1)_{\rm B-L}$ type-III seesaw is the tight, quantitative interplay between the symmetry breaking sector, dark matter and collider observables. The vacuum expectation value $v_{\Phi}$ of the singlet scalar $\Phi$ simultaneously sets the dark matter mass $M_{\chi}$ and controls the dynamics of the U(1)$_{\rm B-L}$ phase transition, thereby fixing the characteristic frequency and amplitude of the resulting stochastic gravitational-wave spectrum. The same parameters, in particular the gauge coupling $g_{\rm B-L}$ and the $Z_{\rm BL}$ mass not only determine the DM relic abundance and detection prospects, but also govern the $s$-channel enhancement of fermion-triplet pair production at colliders. Crucially, the charged component of the triplet, $\Sigma^{\pm}$ can decay via $\Sigma^{\pm}\to\Sigma^{0}\pi^{\pm}$ which arises due to small electroweak corrections, typically $M_{\Sigma^\pm}-M_{\Sigma^0}\sim 166\,$MeV.   This results into  disappearing charged tracks (or displaced vertices) of $\Sigma^\pm$. We demonstrate that in a certain parameter space the above decay channel contribute almost 98\% even though there exist other decay channels of $\Sigma^\pm$ such as $\Sigma^{\pm}\to W^{\pm}\nu,\;Z\ell^{\pm}$, $h\ell^{\pm}$, $\Sigma^0e\nu_e$, $\Sigma^0\mu\nu_\mu$. We show that the above mentioned feature is correlated to the neutrino-oscillation parameters, most notably the lightest neutrino mass $m_{1}$. Thus multi-observable dependence of our scenario yields a concrete complementarity between gravitational wave, dark matter, collider and low-energy neutrino probes, making the scenario predictive and experimentally verifiable in ways that distinguish it from more conventional seesaw+DM models.

The rest of this paper is organized as follows: In Section~\ref{sec:model}, we provide a detailed description of our model, including the particle content, gauge symmetries, and the Lagrangian. Section~\ref{sec:neutrinomass} is dedicated to the neutrino mass generation mechanism through the type-III seesaw in our $U(1)_{\rm B-L}$ framework. In Section~\ref{sec:dm}, we explore the DM phenomenology, discussing the relic density calculations. Section~\ref{sec:det} focuses on the direct and indirect detection prospects and constraints from current experiments on the collider phenomenology of our model, with particular emphasis on the production and decay of triplet fermions and the $\rm B-L$ gauge boson. We also discuss the potential for observing disappearing track signatures in future collider experiments. In Section \ref{sec:gw}, we discuss the gravitational wave signature of the first order phase transition that occurs during the B-L symmetry breaking, providing a complementary probe and  finally conclude in Section~\ref{sec:concl}.

\section{Model}\label{sec:model}

Our model extends the SM gauge group to $SU(3)_C \otimes SU(2)_L \otimes U(1)_Y \otimes U(1)_{\rm B-L}$ by incorporating an additional $U(1)_{\rm B-L}$ symmetry. With this extension, we aim to address two fundamental issues: the origin of neutrino masses and the nature of dark matter (DM) in the canonical type-III seesaw framework. The minimal particle content for a type-III seesaw model includes three $SU(2)_L$ fermion triplets ($\Sigma_{R_1}, \Sigma_{R_2}, \Sigma_{R_3}$) in addition to the SM particles.

Here, it is worth mentioning that, in the case of type-I seesaw model with $U(1)_{\rm B-L}$ symmetry, the $\rm B-L$ anomalies {\it i.e.},
\begin{equation}
\mathcal{A}_1^{\rm SM}[U(1)_{\rm B-L}^3]=-3 ~~; ~~ \mathcal{A}_2^{\rm SM}[(grav.)^2\times U(1)_{\rm B-L}]=-3, 
\end{equation}are automatically canceled by the introduction of three right-handed neutrinos with the sum of their $\rm B-L$ charges equal to $-3$. However, this automatic cancellation does not occur in the type-III seesaw scenario. The key difference lies in the fact that the fermion triplets in type-III seesaw are not SM gauge singlets, unlike the right-handed neutrinos in type-I seesaw.

 	\begin{table}[h]
    \resizebox{8cm}{!}{
	\begin{tabular}{|c|c|c|c|}
		\hline \multicolumn{2}{|c}{Fields}&  \multicolumn{1}{|c|}{ $\underbrace{ SU(3)_{ \rm C} \otimes SU(2)_{\rm L} \otimes U(1)_{\rm Y}}$ $\otimes   U(1)_{\rm B-L} $} \\ \hline
     	&  $\Sigma_{R_1}$&  ~~1 ~~~~~~~~~~~3~~~~~~~~~~0~~~~~~~~~ -1 \\ [0.5em] \cline{2-3}
		& $\Sigma_{R_2}$ &~1 ~~~~~~~~~~~3~~~~~~~~~~~0~~~~~~~~~-1 \\ [0.5em]
		\cline{2-3}
		{Fermions} 	&  $\Sigma_{R_3}$ &~1 ~~~~~~~~~~~3~~~~~~~~~~~0~~~~~~~~~2 \\ [0.5em]
		\cline{2-3}
        &  $\chi_{_L}$ &~~~~~1 ~~~~~~~~~~~1~~~~~~~~~~0~~~~~~~~$\frac{9+\sqrt{57}}{6}$ \\ [0.5em]
		\cline{2-3}
		&  $\chi_{_R}$ &~~~~~~1 ~~~~~~~~~~~1~~~~~~~~~~0~~~~~~~~$\frac{-9+\sqrt{57}}{6}$ \\  [0.5em]
		\cline{1-3}
		&$\Phi_1$ & ~~~1 ~~~~~~~~~~1~~~~~~~~~~0~~~~~~~~~~ 2 \\
		{Scalar} & $\Phi_2$ & ~~~1 ~~~~~~~~~~1~~~~~~~~~~0~~~~~~~~~~ 4 \\
    	&$\Phi_3$ &~~~1 ~~~~~~~~~~1~~~~~~~~~~0~~~~~~~~~~ -1 \\
    	&$\Phi$ &~~~1 ~~~~~~~~~~1~~~~~~~~~~0~~~~~~~~~~ 3\\		
		\hline		

	\end{tabular}
}
	\caption{Charge assignment of BSM fields under the gauge group $\mathcal{G} \equiv \mathcal{G}_{\rm SM} \otimes U(1)_{\rm B-L} $,  where $\mathcal{G}_{\rm SM}\equiv SU(3)_c \otimes SU(2)_L \otimes U(1)_Y$. }
	\label{tab:tab1}
\end{table}

To ensure anomaly cancellation in our $\rm B-L$ symmetric type-III seesaw model, we must consider the three non-trivial anomaly conditions:
i) $SU(2)_L^2 \times U(1)_{\rm B-L}$, 
ii) $U(1)_{\rm B-L}^3$ and 
iii) $(grav.)^2\times U(1)_{\rm B-L}$. 
Considering the $U(1)_{\rm B-L}$ charges of $\Sigma_{R_1},~ \Sigma_{R_2}$, and $ \Sigma_{R_3} $ as $x_1$, $x_2$, and $x_3$, the three anomalies become,
\begin{equation}
\mathcal{A}_1^{\rm SM+BSM}[SU(2)_L^2 \times U(1)_{\rm B-L}]=-3(x_1+x_2+x_3)
\label{Eq:cond3}	
\end{equation}
\begin{equation}
\mathcal{A}_2^{\rm SM+BSM}[U(1)_{\rm B-L}^3]=-3-3(x_1^3+x_2^3+x_3^3)
\label{Eq:cond1}
\end{equation}
\begin{equation}
\mathcal{A}_3^{\rm SM+BSM}[(grav.)^2\times U(1)_{\rm B-L}]=-3-3(x_1+x_2+x_3)
\label{Eq:cond2}
\end{equation}

Let us first address the $SU(2)_L^2 \times U(1)_{\rm B-L}$ anomaly. This can be canceled by assigning $\rm B-L$ charges of $-1, -1,$ and $2$ to the three fermion triplets {\it i.e.} $x_1=-1, x_2=-1, x_3=2$ respectively. 

However, canceling this anomaly alone is not sufficient. The other two anomalies, $U(1)_{\rm B-L}^3$ and $(grav.)^2\times U(1)_{\rm B-L}$, remain non-zero:

\begin{eqnarray}
\mathcal{A}_2^{\rm SM+BSM}[U(1)_{\rm B-L}^3]=-21 \nonumber \\  \mathcal{A}_3^{\rm SM+BSM}[(grav.)^2\times U(1)_{\rm B-L}]=-3 \label{Eq:con4}
\end{eqnarray}

To cancel these remaining anomalies, we introduce two new chiral fermions, $\chi_L$ and $\chi_R$, with specific $\rm B-L$ charges. The $\rm B-L$ charges of these fermions are calculated to be $(9+\sqrt{57})/6$ and $(-9+\sqrt{57})/6$, respectively\footnote{Several explicit models in the literature utilize irrational or fractional $U(1)$ charges in anomaly-free gauge theories. Some illustrative examples include~\cite{Altmannshofer:2019xda, Kanemura:2014rpa,Han:2017ars,Berbig:2022nre,Nanda:2017bmi,Gu:2019ird,Gu:2019gzy}. It is worth noting that such irrational $U(1)$ charges cannot come from a unified gauge theory but remain perfectly consistent in a stand-alone $U(1)$ extension if anomaly cancellation is maintained~\cite{Altmannshofer:2019xda}.}, ensuring that all three anomaly conditions are simultaneously satisfied. $\chi_L,\chi_R$ constitute a Dirac fermion $\chi$, which serves as the DM candidate in our setup.

In addition to these fermions, we introduce four new scalar fields: $\Phi_1, \Phi_2, \Phi_3$, and $\Phi$. These scalars play crucial roles in our model:
 $\Phi_1, \Phi_2$, and $\Phi_3$ are necessary for generating masses for the fermion triplets after $U(1)_{\rm B-L}$ symmetry breaking and $\Phi$ is introduced to provide mass to the DM candidate.
These additional scalar fields, with their specific $U(1)_{\rm B-L}$ charges, allow us to construct a rich phenomenology that connects neutrino mass generation and DM. The particle content and their charge assignments are listed in Table \ref{tab:tab1} for quick reference. \\

\noindent\underline{\bf Scalar Sector:}\\

The scalar sector of our model consists of the SM Higgs doublet $H$ and four additional complex scalar singlets $\Phi_1$, $\Phi_2$, $\Phi_3$, and $\Phi$.
The scalar Lagrangian of the model consistent with the imposed symmetry can be written as:

\begin{eqnarray}
\mathcal{L}_{scalar}&=&(D_\mu H)^\dagger(D_\mu H)+\sum_{i=1}^{3}(\mathcal{D}_\mu\Phi_i)^\dagger(\mathcal{D}_\mu\Phi_i)\nonumber\\&+&(\mathcal{D}_\mu\Phi)^\dagger(\mathcal{D}_\mu\Phi)- V(H,\Phi_i,\Phi)
\end{eqnarray}

where $D_\mu$ and $\mathcal{D}_\mu$ are the covariant derivatives for the SM and $U(1)_{\rm B-L}$ gauge groups, respectively:
\begin{eqnarray}
D_\mu=\partial_\mu+i\frac{g}{2}\sigma_aW^a_\mu+i\frac{g^\prime}{2}B_\mu\\
\mathcal{D}_\mu=\partial_\mu+ig_{_{\rm BL}}Q_{\rm BL}(Z_{\rm BL})_\mu,
\end{eqnarray}
with $Q_{\rm BL}$ being the $U(1)_{\rm B-L}$ charge of the corresponding field.

The scalar potential $V(H,\Phi_i,\Phi)$ includes mass terms, self-interactions, and cross-couplings between different scalar fields:

\begin{eqnarray}
    V(H,\Phi_i,\Phi)&=&-\mu^2_H(H^\dagger H)+\lambda_H(H^\dagger H)^2+\sum_{i=1}^{3}\big(-\mu^2_{\Phi_i}(\Phi_i^\dagger\Phi_i)\nonumber\\&&+\lambda_{\Phi_i}(\Phi_i^\dagger\Phi_i)^2\big)+(\mu_1 \Phi_1\Phi_1\Phi_2^\dagger+\mu_2\Phi_3\Phi_3\Phi_1\nonumber\\&&+\lambda_1\Phi_3\Phi_3\Phi_2\Phi_1^\dagger)+\sum_{i=1}^{3}\lambda_{H\Phi_i}(H^\dagger H)(\Phi_i^\dagger\Phi_i)\nonumber\\&&+\big(-\mu^2_{\Phi}(\Phi^\dagger\Phi)+\lambda_{\Phi}(\Phi^\dagger\Phi)^2\big)\nonumber\\&&+\lambda_{H\Phi}(H^\dagger H)(\Phi^\dagger\Phi)+\mu_3\Phi_1^\dagger\Phi_3\Phi\nonumber\\&&+\mu_4\Phi_2\Phi_3\Phi^\dagger.
\end{eqnarray}

Here, it is worth mentioning that for simplicity, we assume certain mass hierarchy among the BSM scalars. We assume $\Phi$ to be the lightest among the BSM scalars, while the masses of $\Phi_1$, $\Phi_2$, and $\Phi_3$ have masses much higher than $\Phi$ such that they do not play any crucial role in the DM phenomenology. To simplify the analysis, we decouple the light scalars from the heavy
scalars by considering all the corresponding quartic couplings to be negligible. 

After spontaneous symmetry breaking, we parameterize the neutral components of the low-energy scalars as:
\begin{eqnarray}
\langle H \rangle=\begin{pmatrix}
    0\\\frac{v_h+h}{\sqrt{2}}
\end{pmatrix},\langle\Phi\rangle=\frac{\phi+v_\phi}{\sqrt{2}},
\end{eqnarray}
where $v_h$ is the vev of SM Higgs and $v_\phi$ is the vev of $\Phi$.
The presence of $\lambda_{H\Phi}$ term induces mixing between $H-\Phi$ and gives mass eigenstates as $h_1,h_2$. The mass-squared matrix can be written as
\begin{eqnarray}
\mathcal{M}^2_{h\phi}=\begin{pmatrix}
    2v_h^2\lambda_H&v_hv_\phi\lambda_{H\Phi}\\v_hv_\phi\lambda_{H\Phi}&2v_\phi^2\lambda_\Phi\,,
\end{pmatrix}
\end{eqnarray}
and thus, the mixing angle is given by
\begin{eqnarray}
\tan(\theta_{h\phi})=\frac{v_hv_\phi\lambda_{H\Phi}}{v_h^2\lambda_H-v_\phi^2\lambda_\Phi}\,.
\end{eqnarray}
The scalar couplings can be expressed in terms of the masses and mixing angles as
\begin{eqnarray}
\lambda_H=\frac{\cos^2(\theta_{h\phi})m^2_{h_1}+\sin^2(\theta_{h\phi})m^2_{h_2}}{2v_h^2}
\end{eqnarray}
\begin{eqnarray}
\lambda_\Phi=\frac{\cos^2(\theta_{h\phi})m^2_{h_2}+\sin^2(\theta_{h\phi})m^2_{h_1}}{2v_\phi^2}\,,
\end{eqnarray}
\begin{eqnarray}
\lambda_{H\Phi}=\sin(2\theta_{h\phi})\frac{m^2_{h_1}-m^2_{h_2}}{2v_h v_\Phi}\,.
\end{eqnarray}

After the breaking of $U(1)_{\rm B-L}$ symmetry, $Z_{{\rm BL}}$ boson acquires mass through the vevs of the scalars charged under the $U(1)_{\rm B-L}$, and is given by:
\begin{eqnarray}
M_{Z_{\rm BL}}=\sqrt{4 u^2_1+16 u^2_2+ u^2_3+9v_\phi^2}~g_{_{\rm BL}}.
\end{eqnarray}

where $\langle \Phi_i \rangle = u_i $. 
Assuming  $u_1 \simeq u_2 \simeq u_3 \simeq u\,$, we can write:
\begin{eqnarray}
M_{Z_{\rm BL}}=\sqrt{21 u^2+9v_\phi^2}~g_{_{\rm BL}}.\label{eq:mzbl}
\end{eqnarray}

\section{Neutrino mass}\label{sec:neutrinomass}
In our model, neutrino masses are generated through the type-III seesaw mechanism, which is embedded within the $U(1)_{\rm B-L}$ framework. As mentioned earlier, the type-III seesaw mechanism introduces $SU(2)_L$ fermion triplets $\Sigma_{R_i}\sim (i = 1, 2, 3)$ with zero hypercharge. These triplets, in their adjoint representation, can be written as:
\begin{eqnarray*}
 \Sigma_{R_i}=\begin{pmatrix}
    \Sigma^0_i/\sqrt{2} & \Sigma^+_i \\
\Sigma^-_i &  -\Sigma^0_i/\sqrt{2}
 \end{pmatrix}   
\end{eqnarray*}

The relevant Lagrangian for neutrino mass generation is:
\begin{eqnarray}
    \mathcal{L}_\nu \supset y_{\alpha i} \overline{L_\alpha}\tilde{H}\Sigma_{R_i} -\frac{1}{2} M_{\Sigma_i} {\rm Tr}[\overline{\Sigma^c_{R_i}}\Sigma_{R_i}]+h.c.
\end{eqnarray}
where $L_\alpha$ is the left-handed lepton doublet, $H$ is the SM Higgs doublet, $Y_{\alpha i}$ are the Yukawa couplings, and $M_{\Sigma_i}$ are the masses of the fermion triplets.

After electroweak symmetry breaking, the neutrino mass matrix in the basis $(\nu^c_L, \Sigma^0)$ takes the form:
\begin{eqnarray}
    M_\nu = \begin{pmatrix}
    0 & M_D \\
M^T_D &  M_\Sigma
 \end{pmatrix}   
\end{eqnarray}\,

We can diagonalize this matrix using the seesaw approximation, yielding the light neutrino mass matrix:
\begin{equation}
M_\nu=-M_D M_\Sigma^{-1}M_D^T
\end{equation}

In our $U(1)_{\rm B-L}$ extended model, the masses of the fermion triplets are generated through the spontaneous breaking of the B-L symmetry. The relevant Yukawa interactions are:
\begin{eqnarray}
\mathcal{L}_{\Sigma.}&=&y_{_{\Sigma_{11}}}\overline{\Sigma_{R_1}^c}\Sigma_{R_1}\Phi_1+y_{_{\Sigma_{22}}}\overline{\Sigma_{R_2}^c}\Sigma_{R_2}\Phi_1+y_{_{\Sigma_{33}}}\overline{\Sigma_{R_3}^c}\Sigma_{R_3}\Phi_2\nonumber\\&+&y_{_{\Sigma_{12}}}(\overline{\Sigma_{R_1}^c}\Sigma_{R_2}+\overline{\Sigma_{R_2}^c}\Sigma_{R_1})\Phi_1 \nonumber\\&+&y_{_{\Sigma_{13}}}(\overline{\Sigma_{R_1}^c}\Sigma_{R_3}+\overline{\Sigma_{R_3}^c}\Sigma_{R_1})\Phi_3\nonumber\\&+&y_{_{\Sigma_{23}}}(\overline{\Sigma_{R_2}^c}\Sigma_{R_3}+\overline{\Sigma_{R_3}^c}\Sigma_{R_2})\Phi_3 \label{Eq:majorana}
\end{eqnarray}

Thus, after $U(1)_{\rm B-L}$ symmetry breaking, these terms generate the following mass matrix for the triplet fermions:
 \begin{equation}
M_\Sigma=\begin{pmatrix}
y_{_{\Sigma_{11}}}u & y_{_{\Sigma_{12}}}u & y_{_{\Sigma_{13}}}u\\
y_{_{\Sigma_{12}}}u & y_{_{\Sigma_{22}}}u & y_{_{\Sigma_{23}}}u\\
y_{_{\Sigma_{13}}}u & y_{_{\Sigma_{23}}}u & y_{_{\Sigma_{33}}}u
\end{pmatrix}\,,
\end{equation}	
and similarly, after the electroweak symmetry breaking, the Dirac mass matrix takes the form:
\begin{equation}
 M_D=\begin{pmatrix}
 y_{_{11}}v_h & y_{_{12}}v_h & 0\\
 y_{_{21}}v_h & y_{_{22}}v_h & 0\\
 y_{_{31}}v_h & y_{_{32}}v_h & 0
 \end{pmatrix}\,.
 \end{equation}	

The structure of these mass matrices reveals that our model is flexible enough to accommodate both normal and inverted neutrino mass hierarchies while remaining consistent with current neutrino oscillation data. 

\section{Dark matter}\label{sec:dm}	

In our model, the Dirac fermion $\chi$, formed by the combination of $\chi_L$ and $\chi_R$, emerges as a natural dark matter (DM) candidate. The stability of $\chi$ as a DM candidate is ensured by a residual $Z_2$ symmetry, which remains after the spontaneous breaking of the $U(1)_{\rm B-L}$ symmetry. This symmetry forbids the decay of $\chi$ into SM particles, making it a viable DM candidate.

The relevant Lagrangian for the SM singlets introduced for anomaly cancellation is given by:
\begin{eqnarray}
\mathcal{L}_{\chi}&=& \overline{\chi_{_L}}i\gamma^\mu \mathcal{D}_\mu\chi_{_L}+ \overline{\chi_{_R}}i\gamma^\mu \mathcal{D}_\mu\chi_{_R}+y_{\chi}\overline{\chi_{_L}}\chi_{_R}\Phi\nonumber\\&+&y_{\chi}\overline{\chi_{_R}}\chi_{_L}\Phi^\dagger.\label{eq:Ldm1}
\end{eqnarray}
Define a Dirac fermion $\chi$ as
$\chi=\chi_{_L}+\chi_{_R}$. Then Eq \ref{eq:Ldm1} can be rewritten as,
\begin{eqnarray}
\mathcal{L}_{\chi}&=&\bar{\chi}i\gamma^\mu\partial_\mu\chi-g_{_{\rm BL}}\bar{\chi}\gamma^\mu (Z_{\rm BL})_\mu\left( \frac{-9+\sqrt{57}}{6} \right)\chi\nonumber\\&-&y_\chi\bar{\chi}P_R\chi\Phi-y_\chi\bar{\chi}P_L\chi\Phi^\dagger,
\end{eqnarray}
where $P_{L/R}$ are the projection operators defined as $P_{L/R}=({1\mp\gamma_5})/{2}$.

When $\Phi$ obtains a vev $v_\phi$, and the $U(1)_{\rm B-L}$ symmetry gets broken spontaneously, $\chi$ gets a Dirac mass as:
\begin{eqnarray}
M_{\rm DM}\equiv M_{\rm \chi}=y_\chi v_\phi\,. \label{eq:mdm}
\end{eqnarray}	

The primary interactions of $\chi$ with the SM particles occur through the $U(1)_{\rm B-L}$ gauge boson $Z_{\rm BL}$ and 
the scalar mediator $h_2$ (the physical state resulting from the mixing of $\Phi$ and $H$).
These interactions play crucial roles in determining the DM relic density and its detection prospects.
The relic abundance of $\chi$ is primarily determined by its annihilation into SM particles through s-channel $Z_{\rm BL}$ and $h_2$ exchange. The dominant annihilation channels include: $\chi\chi\rightarrow ff,ZZ,W^+W^- $, $\chi\chi\rightarrow h_{1,2}h_{1,2}$, $\chi\chi\rightarrow Z_{\rm BL}Z_{\rm BL}$, $\chi\chi\rightarrow Z_{\rm BL}h_{1,2}$ and $\chi\chi\rightarrow \Sigma^+\Sigma^-$, if kinematically allowed.

The DM relic density can be calculated by solving the Boltzmann equation
\begin{eqnarray}
\frac{dY_\chi}{dx}=-\frac{sx}{\mathcal{H}(M_{\rm DM})}
\langle\sigma v\rangle_{\chi}\left(Y_\chi^2-(Y^{eq}_{\chi})^2\right)
\end{eqnarray}
where $x=M_{\rm DM}/T$ is the dimensionless parameter to track the evolution of temperature, $Y^{eq}_\chi$ is the equilibrium abundance of $\chi$ {\it i.e.} $Y^{eq}_\chi=0.116\frac{g_\chi}{g_*}x^2K_2(x)$, $K_2(x)$ is the modified Bessel function of second kind. $\mathcal{H}$ is the Hubble parameter $\mathcal{H}(M_{\rm DM})=1.66\sqrt{g_*}{M_{\rm DM}^2}/{M_{pl}}$, $s$ is the entropy density $s=\frac{2\pi^2}{45}g_*M_{\rm DM}^3x^{-3}$ and $\langle\sigma v\rangle_{\chi}$ is the thermally averaged annihilation cross-section. Here, $M_{pl}=1.22\times10^{19}$ GeV is the Planck mass, and $g_*=106.75$ is the relativistic degrees of freedom (d.o.f). 

It is worth mentioning here that we used the package \texttt{micrOMEGAs} \cite{Alguero:2023zol} for computing
annihilation cross-section and relic density, after generating the model files using \texttt{LanHEP} \cite{Semenov:2014rea}.

Before delving deeper into the DM phenomenology, it's crucial to address the constraints on scalar portal couplings arising from Higgs invisible decay searches. In our model, the SM-like Higgs boson $h_1$ can potentially decay into pairs of the lighter scalar $h_2$ ($h_1 \to h_2h_2$) or into DM pairs ($h_1 \to \chi \bar{\chi}$) through scalar mixing. These decay channels contribute to the Higgs invisible width, which is tightly constrained by experimental searches.
The most recent bound on the Higgs invisible decay branching ratio comes from ATLAS collaboration, setting an upper limit of 14.5\% at 95\% confidence level \cite{ATLAS:2022yvh}. This constraint significantly restricts the allowed parameter space for scalar portal couplings and mixing angles as well as DM Yukawa couplings.
In our subsequent analysis, we have carefully accounted for this constraint, ensuring that all considered scenarios comply with the current Higgs invisible decay limit. This consideration is crucial as it directly impacts the viable regions of parameter space for DM phenomenology, particularly for DM masses below $M_{h_1}/2 \simeq 62.5$ GeV.

\subsection{Parameter Space Scan}
To comprehensively understand the DM relic density and elucidate the specific roles of various model parameters in achieving the observed abundance, we conducted extensive analyses and parameter space scans. The key parameters governing
the relic abundance of DM are $M_{\rm DM}$, $y_\chi$, $\sin(\theta_{h\phi})$, $g_{\rm BL}$. Apart from these, other crucial parameters that have a noteworthy effect on DM relic, as well as other phenomenological aspects, are $M_{ Z_{\rm BL}}$ and $M_{h_2}$.
These parameters play crucial roles in determining the strength and nature of DM interactions, influencing annihilation channels, and shaping the viable parameter space that satisfies both the observed relic density and other experimental constraints. 
To systematically explore the parameter space, we follow a structured approach: we treat $M_{\rm DM}$ and $y_{\chi}$ as free parameters and determine the vev of $\Phi$, denoted as $v_\phi$, using Eq.~\ref{eq:mdm}. We assume that the vev of $\Phi_{1,2,3}$ is 100 times that of $\Phi$, i.e., $u = 100 v_\phi$, and subsequently compute $g_{\rm BL}$ using Eq.~\ref{eq:mzbl}.  

Thus, the final set of free parameters for our parameter scan is:  
\begin{equation}
\{M_{\rm DM}, y_\chi, \sin(\theta_{h\phi}), M_{Z_{\rm BL}}, M_{h_2} \}.
\end{equation}
 
\begin{figure}[h]
	\centering	\includegraphics[scale=0.32]{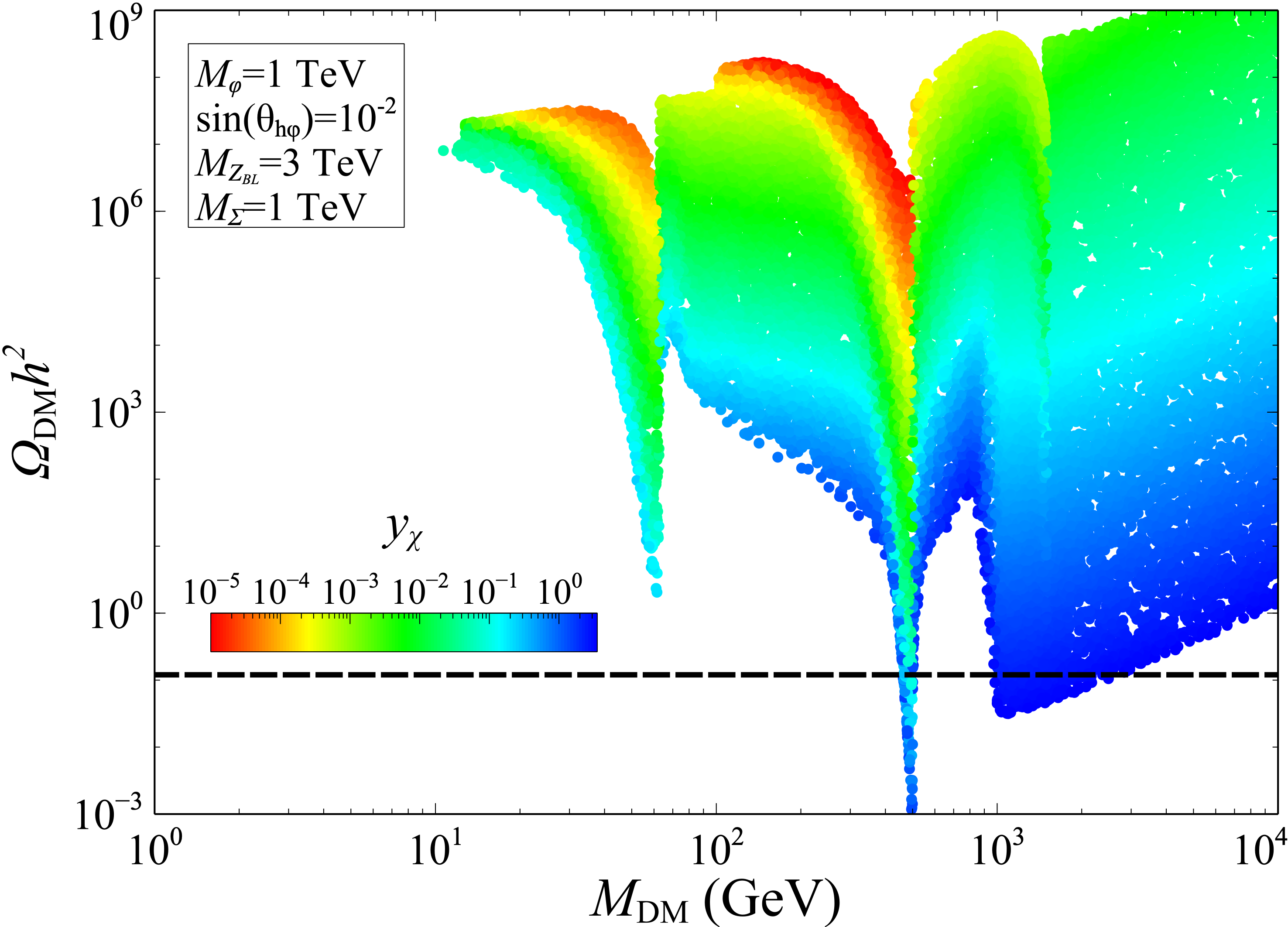}\caption{Relic density as a function of DM mass, with the color bar indicating variations in Yukawa coupling $y_\chi$. Other free parameters are kept fixed as mentioned in the inset of the figure.}\label{fig:scan1}
\end{figure}

Initially, we set the following parameters: the mass of the scalar particle $h_2$ to 1\,TeV, the scalar mixing angle $\sin(\theta_{h\phi})$ to $10^{-2}$, the mass of the $Z_{\rm BL}$ boson to 3\,TeV, the gauge coupling constant $g_{_{\rm BL}}$ to 0.1, and the mass of $\Sigma$ to 1\,TeV. We varied the Yukawa coupling $y_\chi$ within the range $\{10^{-5},\sqrt{4\pi}\}$ and calculated the DM relic density for DM masses spanning from 1\,GeV to 10\,TeV, as depicted in Fig.~\ref{fig:scan1}. {Three resonance peaks are observed at 62.5 GeV, 500 GeV, and 1.5 TeV, corresponding to to the Standard Model Higgs boson ($h_1$), the additional scalar ($h_2$), and the $Z_{\rm BL}$ boson respectively.} When the DM mass exceeds the mass of $h_2$, the $\chi\bar{\chi}\rightarrow h_2h_2$ channel becomes kinematically accessible, leading to an increase in the annihilation cross-section and a consequent decrease in the relic density. Additionally, for DM masses above 2\,TeV, the $\chi\bar{\chi}\rightarrow h_2 Z_{\rm BL}$ channel opens up, and DM predominantly annihilates through this mode even for very small values of $y_\chi$. We can clearly see that with an increase in $y_\chi$, the relic density decreases due to the annihilation cross-section being proportional to $y^2_\chi$. Furthermore, the effect of the DM mass on the relic density is evident when $M_{\rm DM}$ is away from resonances, \textit{i.e.,} for $M_{\rm DM} > 1\,\text{TeV}$. As the DM mass increases, the relic density gradually increases because of the decrease in the annihilation cross-section, which is inversely proportional to the DM mass.

\begin{figure}[h]
	\centering	\includegraphics[scale=0.32]{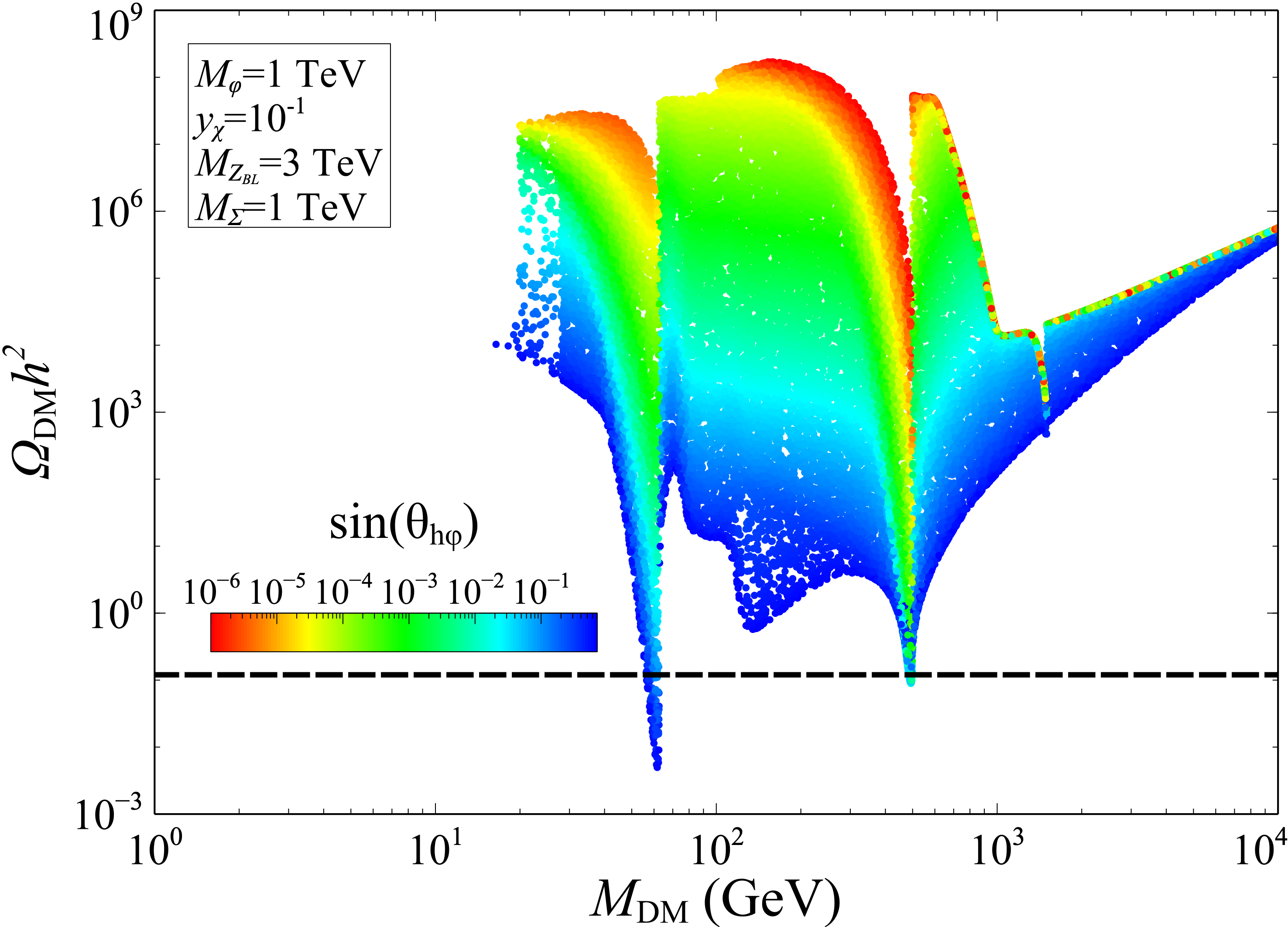}\caption{Dependence of relic density on DM mass. Free parameters are varied as mentioned.}\label{fig:scan2}
\end{figure}

In Fig.~\ref{fig:scan2}, we analyze the impact of the scalar mixing angle, $\sin(\theta_{h\phi})$, on the relic density by varying it randomly within the range $\{10^{-6}, 0.707\}$. The figure illustrates the relic density as a function of DM mass, with the color gradient representing the variation in the mixing angle. Here, we fix the Yukawa coupling $y_\chi$ at 0.1 and maintain the other parameters as specified in Fig. \ref{fig:scan1}. Three resonance peaks are evident, corresponding to the $h_1$, $h_2$, and $Z_{\rm BL}$ mediated s-channel processes. As the value of the mixing angle increases, the relic density decreases. This behavior arises because the annihilation cross-section is proportional to the square of the mixing angle; thus, an increase in the mixing angle enhances the annihilation cross-section, leading to a reduction in relic density.

\begin{figure*}[t]
	\centering
    \begin{tabular}{cc}
	\includegraphics[scale=0.34]{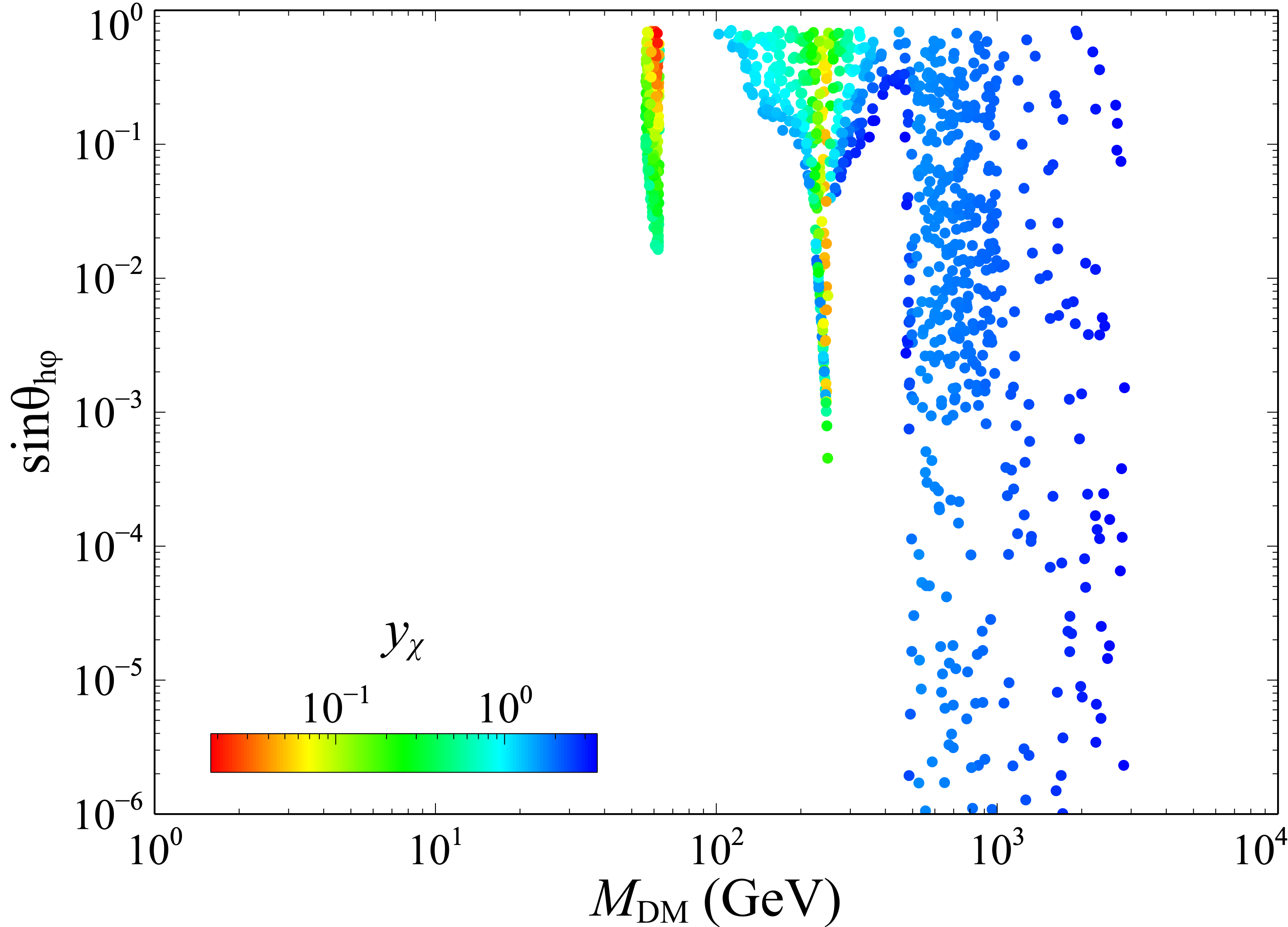}
    \includegraphics[scale=0.34]{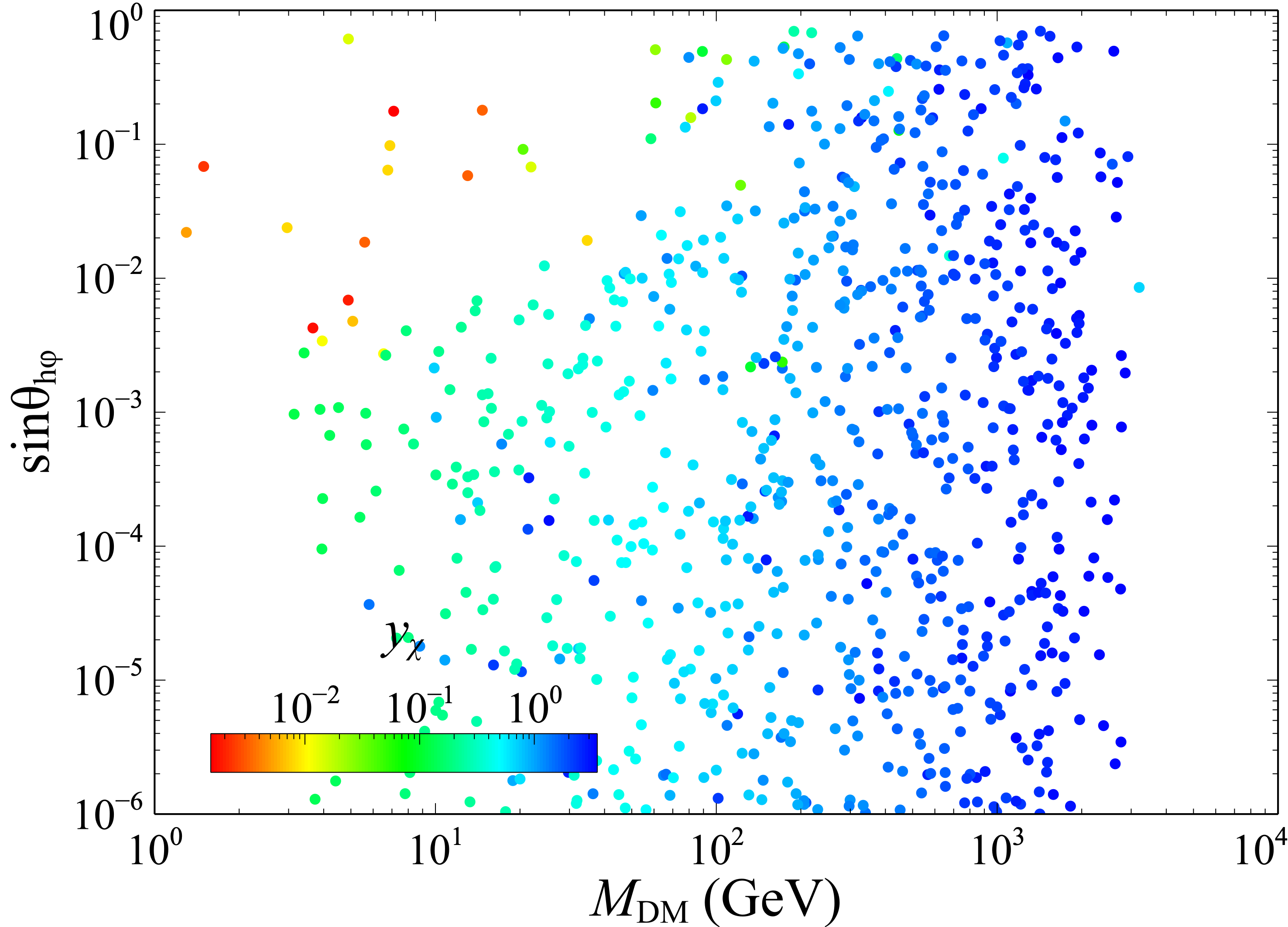}
    \end{tabular}\caption{\textit{left}: Case-I: correct relic points in the plane of $\sin\theta_{h\phi}$ and $M_{\rm DM}$ for fixed $\phi$ mass at 500 GeV. \textit{right}: Case-II: correct relic points in the plane of $\sin\theta_{h\phi}$ and $M_{\rm DM}$.}\label{fig:case1sinvsmdm}
\end{figure*}

To further elucidate the role of beyond Standard Model (BSM) scalar particles in achieving the correct relic density and expanding the viable DM parameter space, we analyze two distinct scenarios:  
\textbf{Case I:} The mass of the scalar $h_2$ is fixed at 500\,GeV.  
\textbf{Case II:} The mass of the scalar $h_2$ is varied randomly over a broad range.  

\subsubsection{Case-I: $M_{h_2}=500$ \rm GeV}\label{subsub:case1}

We perform a comprehensive parameter scan to identify regions that satisfy the correct relic density, varying the parameters within the following ranges:  
\begin{itemize}
    \item $M_{\rm DM} \in [1,10^4]$ GeV,
    \item $y_{\chi} \in [10^{-5},\sqrt{4\pi}]$,
    \item $\sin(\theta_{h\phi}) \in [10^{-6},0.707]$,
    \item $M_{Z_{\rm BL}} \in [1,5]$ TeV.
    \item $M_{h_2} = 500$ GeV,
\end{itemize}  

The points satisfying the relic density constraint are shown in the \textit{left} panel of Fig.~\ref{fig:case1sinvsmdm}, plotted in the $\sin \theta_{h\phi}$–$M_{\rm DM}$ plane. Notably, two dips appear at around 62.5 GeV and 250 GeV, corresponding to $s$-channel processes mediated by the SM Higgs ($h_1$) and the BSM scalar ($h_2$), respectively. At these resonances, the correct relic density is achieved with smaller couplings due to enhanced annihilation cross-sections.  

For $M_{\rm DM} > 500$ GeV, relic density is satisfied for smaller $\sin\theta_{h\phi}$ but with larger Yukawa couplings. These points correspond to the process $\chi\chi \rightarrow h_2 h_2$. Additionally, for larger mixing angles in this mass range, the process $\chi\chi \rightarrow h_1 h_2$ also contributes to determining the relic abundance. When $M_{\rm DM} \gtrsim 750$ GeV, the annihilation channel $\chi\chi \rightarrow h_2 Z_{\rm BL}$ becomes kinematically accessible, further affecting the relic density.  

As the DM mass increases, achieving the correct relic abundance requires progressively larger Yukawa couplings. This arises because higher DM masses lead to a reduced annihilation cross-section, necessitating an enhancement via stronger couplings to maintain the required depletion rate.  

We observe that the correct DM relic density can only be obtained for $M_{\rm DM} > 55$ GeV. However, as we will discuss in the next section, direct detection constraints impose an upper limit on the mixing angle, excluding $M_{\rm DM} < 224$ GeV. Thus to explore viable parameter space for smaller DM masses, we extend our analysis to Case-II, where the scalar mass is varied both above and below the DM mass.

\subsubsection{Case-II: $M_{h_2}\in[1,10^4]$ \rm GeV}\label{subsub:case2}

In this scenario, we allow the mass of the scalar $h_2$ to vary over a broad range, $M_{h_2} \in [1,10^4]$ GeV. The resulting parameter space satisfying the correct relic density is shown in the \textit{right} panel of Fig.~\ref{fig:case1sinvsmdm}, plotted in the $\sin\theta_{h\phi}$–$M_{\rm DM}$ plane.  

For smaller DM masses, the $t$-channel annihilation process $\chi\chi \rightarrow h_2 h_2$ becomes accessible, significantly increasing the cross-section and thereby reducing the relic abundance to the observed level. Consequently, this scenario allows for viable DM candidates in the lower mass range.  

When $M_{h_2}$ exceeds the DM mass, resonant annihilation via $h_2$ plays a crucial role in achieving the correct relic density, particularly for parameter points where $M_{\rm DM} \sim M_{h_2}/2$. This resonance effect, in conjunction with the $Z_{\rm BL}$ resonance, facilitates the depletion of DM.  

Overall, while the general features of the relic density for this case remain similar to those in Case-I, the variation in $h_2$ mass enhances the viable DM parameter space, providing additional flexibility in satisfying both relic density and other experimental constraints.

\section{Detection Prospects}\label{sec:det}
\subsection{Direct Detection}

\begin{figure*}[t]
	\centering
	\includegraphics[scale=0.34]{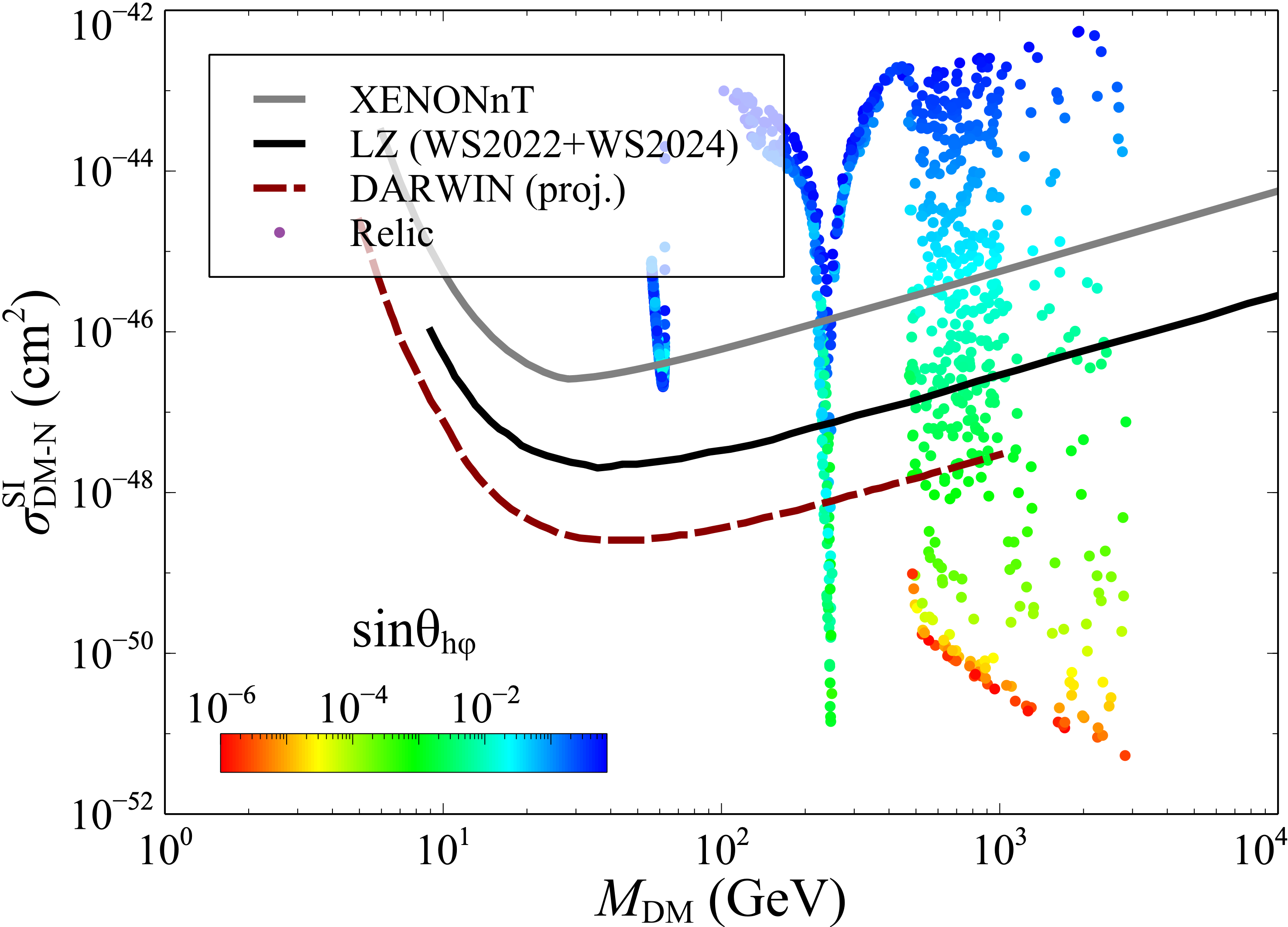}
    \includegraphics[scale=0.34]{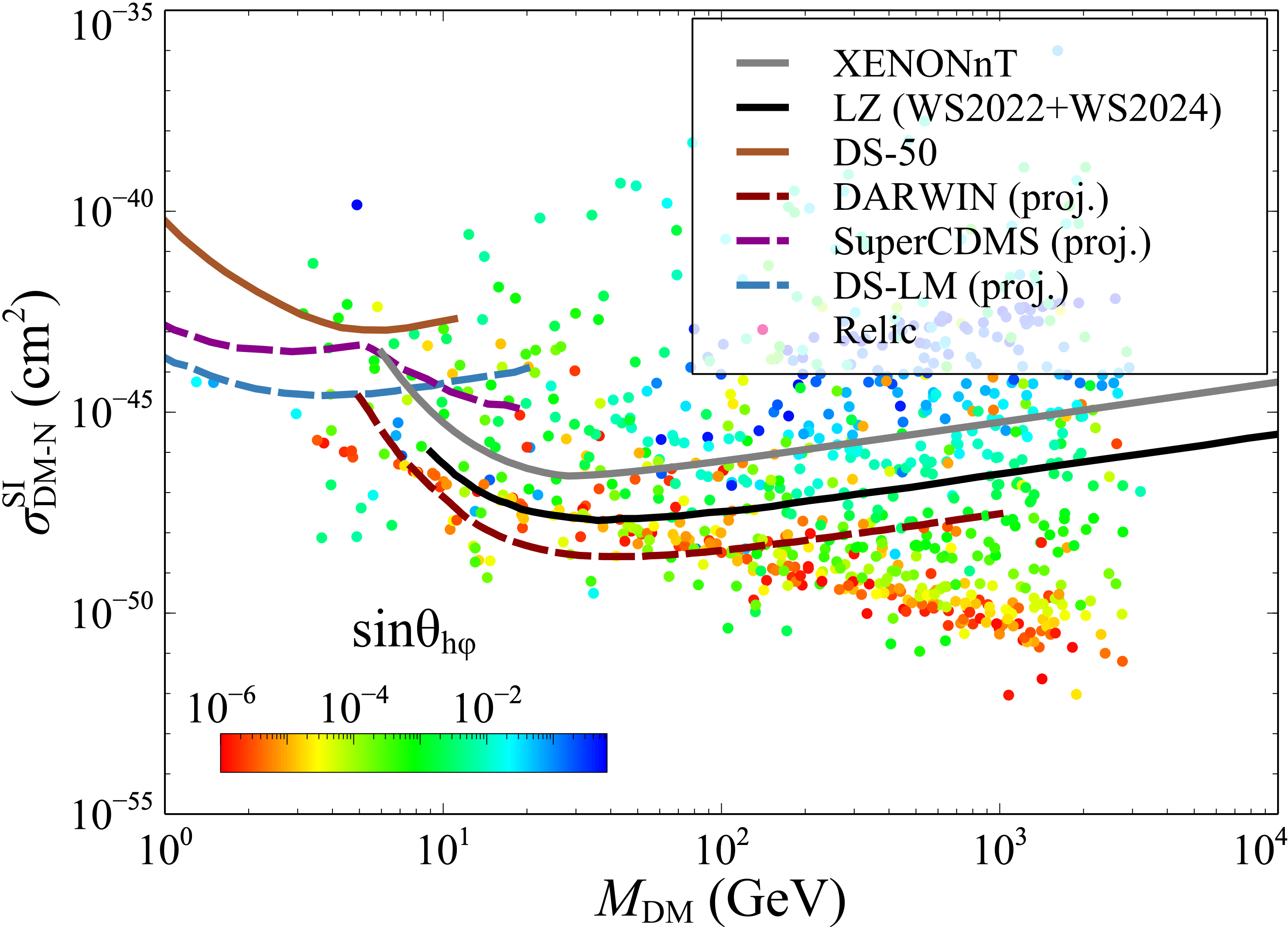}
    \caption{Case-I (\textit{left}), Case-II (\textit{right}): Spin-independent direct detection cross-section as a function of DM mass. {The present constraints and future sensitivities from the different direct detection experiments are shown with different colored lines as mentioned in the legend, see text for details.}}\label{fig:ddxseccase1}
\end{figure*}

\begin{figure}[h]
    \centering   \includegraphics[scale=0.25]{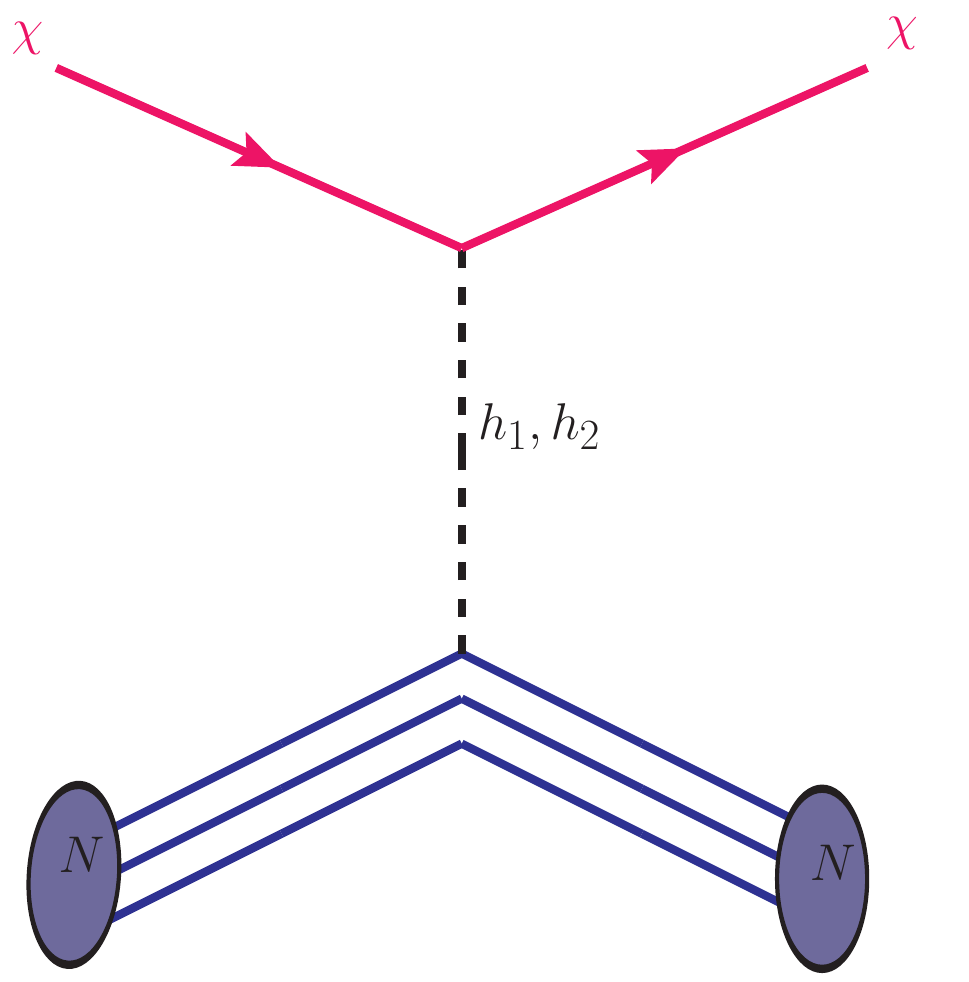}   \includegraphics[scale=0.25]{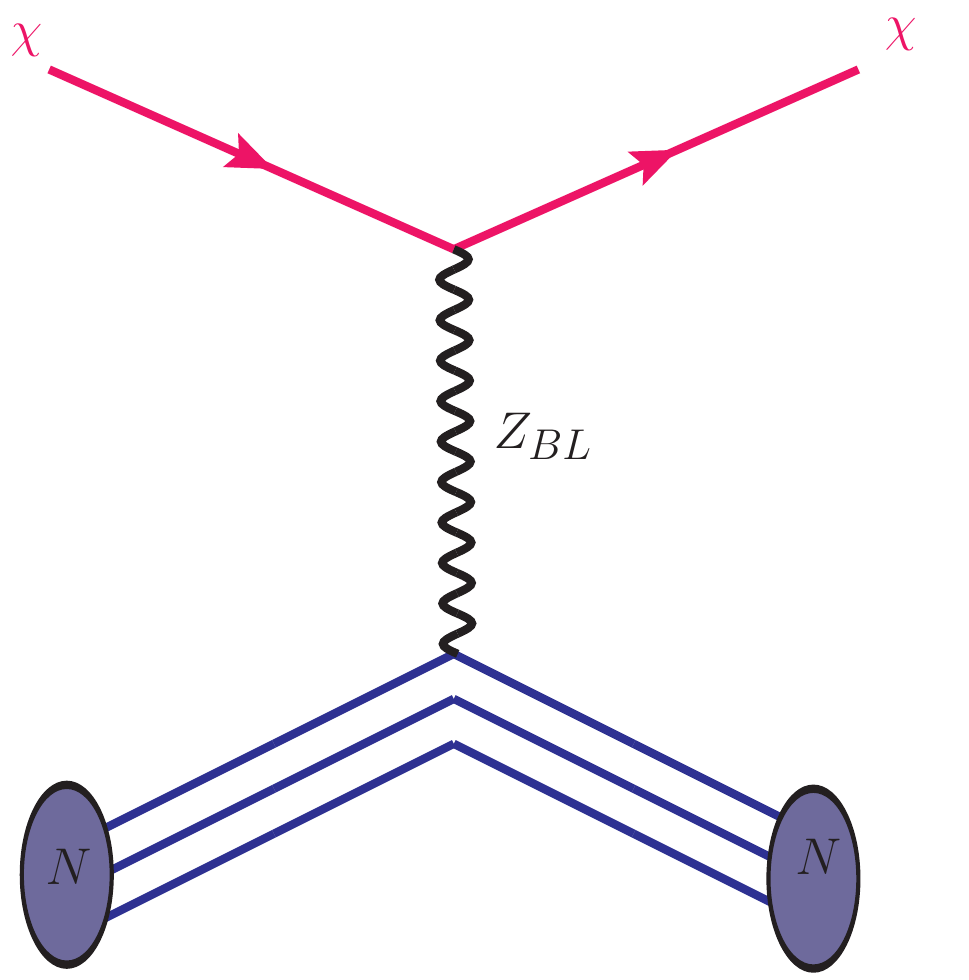}
    \caption{[\textit{left}]: Scalar mediated DM-N scattering, [\textit{right}]: $Z_{\rm BL}$ mediated DM-N scattering}
    \label{fig:xNtoxN}
\end{figure}

In this framework, DM can scatter off the nucleons at terrestrial detectors through both scalar mediation and $U(1)_{\rm B-L}$ vector boson mediation, as illustrated in Fig.~\ref{fig:xNtoxN}.

The spin-independent (SI) DM-nucleon cross-section via scalar mediation is given by \cite{Ellis:2008hf}:
\begin{equation}
\sigma^{\rm SI}_{\rm scalar} = \frac{\mu_r^2}{\pi A^2} [\mathcal{M}]^2,
\end{equation}
where $\mu_r = \frac{M_{\rm DM} m_n}{M_{\rm DM} + m_n}$ is the reduced mass, $m_n$ is the nucleon (proton or neutron) mass, and $\mathcal{M}$ is the amplitude corresponding to the scalar-mediated diagram shown in Fig.~\ref{fig:xNtoxN} [\textit{left}]. This amplitude is expressed as:
\begin{eqnarray}
\mathcal{M} = Z f_{p} + (A - Z) f_{n},
\end{eqnarray}
where $A$ is the mass number of the target nucleus, $Z$ is its atomic number, and $f_{p}$ and $f_{n}$ are the interaction strengths of the DM with protons and neutrons, respectively. These interaction strengths are given by:
\begin{equation}
f^i_{p,n} = \sum_{q=u,d,s} f_{T_q}^{p,n} \alpha^i_{q} \frac{m_{p,n}}{m_q} + \frac{2}{27} f_{T_G}^{p,n} \sum_{q=c,t,b} \alpha^i_{q} \frac{m_{p,n}}{m_q},
\end{equation}
where
\begin{eqnarray}
\alpha^1_{q} &=& y_\chi \sin\theta_{h\phi} \times \frac{m_q}{v} \times \bigg(\frac{1}{m^2_{h_1}}\bigg),\nonumber\\\alpha^2_{q} &=& y_\chi \cos\theta_{\phi h} \times \frac{m_q}{v} \times \bigg(\frac{1}{m^2_{h_2}}\bigg).
\end{eqnarray}

The values of $f_{T_q}^{p,n}$ and $f_{T_G}^{p,n}$ can be found in \cite{Ellis:2000ds}.

The SI DM-nucleon cross-section via $Z_{\rm BL}$ mediation is given by \cite{Berlin:2014tja}:
\begin{eqnarray}
\sigma^{{\rm SI}}_{{\rm vector}} &=& \frac{\mu_r^2}{\pi M^4_{Z_{\rm BL}}} g_{_{\rm BL}}^2 \bigg(\frac{9 + \sqrt{57}}{6}\bigg)^2 \bigg[Z(2\tilde{b}_u + \tilde{b}_d) \nonumber \\
&+& (A - Z)(\tilde{b}_u + 2\tilde{b}_d)\bigg]^2,
\end{eqnarray}
where $\tilde{b}_{u/d}$ is the quark-$Z_{\rm BL}$ coupling, which is the same for all quarks and equals $g_{_{\rm BL}}/3$.

Depending on the gauge coupling $g_{\rm BL}$ and the product of the Yukawa coupling $y_\chi$ with the mixing angle $\theta_{\phi h}$, either the scalar-mediated process or the vector boson-mediated process dominates, or both become comparable. We compute the DM-nucleon scattering cross-section using \texttt{micrOMEGAs} for parameter points that satisfy the correct relic density constraint and present the results as a function of DM mass for Cases I (\ref{subsub:case1}) and II  (\ref{subsub:case2}) in Fig.~\ref{fig:ddxseccase1}, shown in the \textit{left} and \textit{right} panels, respectively.  

These points are then compared against the most stringent upper limits on the DM-nucleon scattering cross-section from the LUX-ZEPLIN (LZ) direct detection experiment~\cite{LZ:2024zvo}, as well as XENONnT \cite{XENON:2023cxc} and DS-50 \cite{DarkSide-50:2022qzh}.  

In Case I, aside from the $h_2$ resonance around 250 GeV, the allowed DM mass range extends from 500 GeV to 3 TeV. In Case II, we find a viable parameter space for DM masses ranging from $\mathcal{O}(1)$ GeV up to 3 TeV. In both scenarios, the maximum mixing angle permitted by direct detection constraints is approximately $\sin(\theta_{h\phi}) \sim 0.3$.  

Additionally, we showcase the projected sensitivities of DARWIN~\cite{DARWIN:2016hyl}, SuperCDMS \cite{SuperCDMS:2016wui}, and DS-LM \cite{GlobalArgonDarkMatter:2022ppc}, which has the potential to probe a significant portion of the parameter space in future experiments.

\subsection{Indirect Detection}
\begin{figure*}[t]
	\centering
	\includegraphics[scale=0.34]{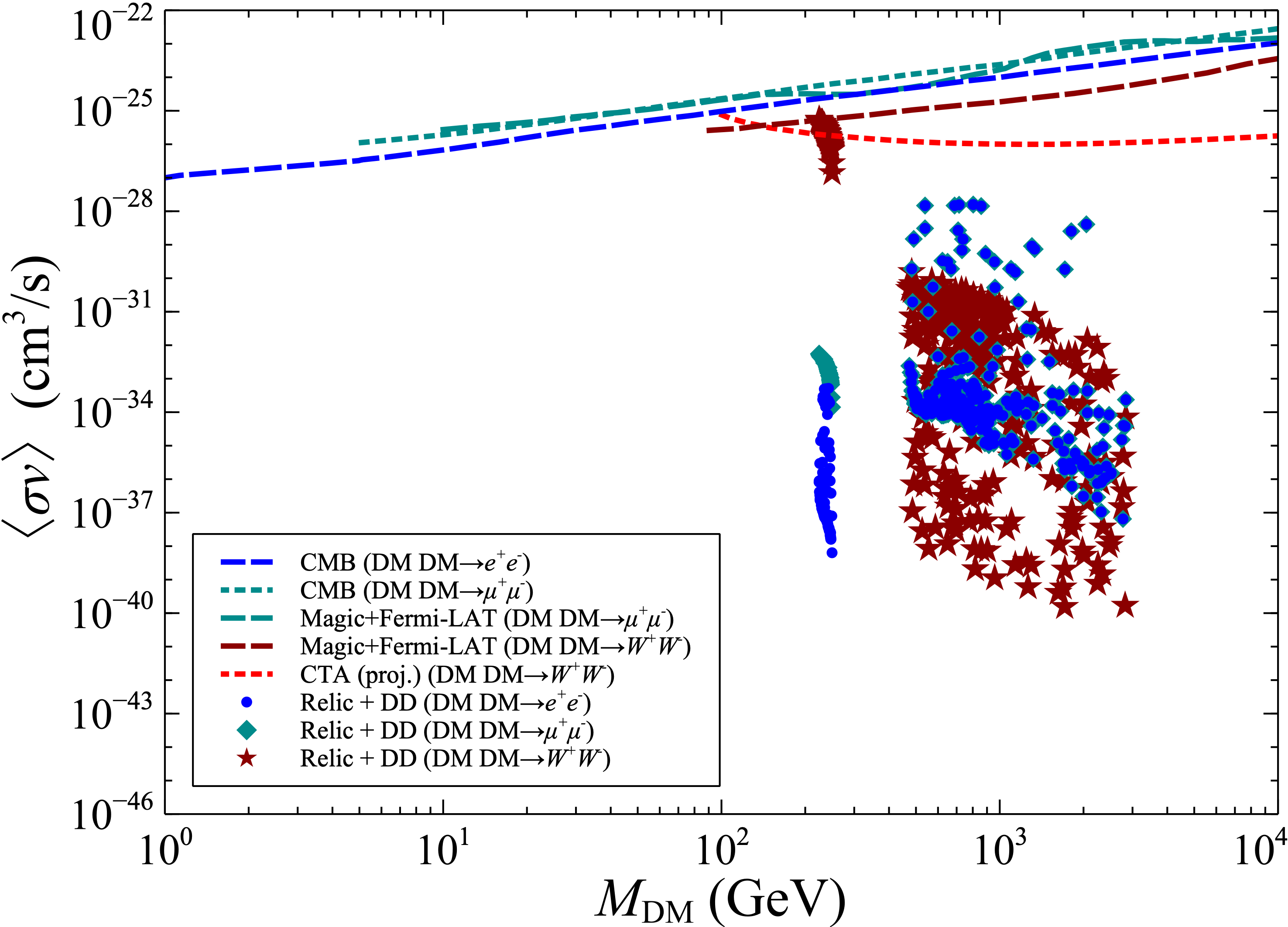}
    \includegraphics[scale=0.34]{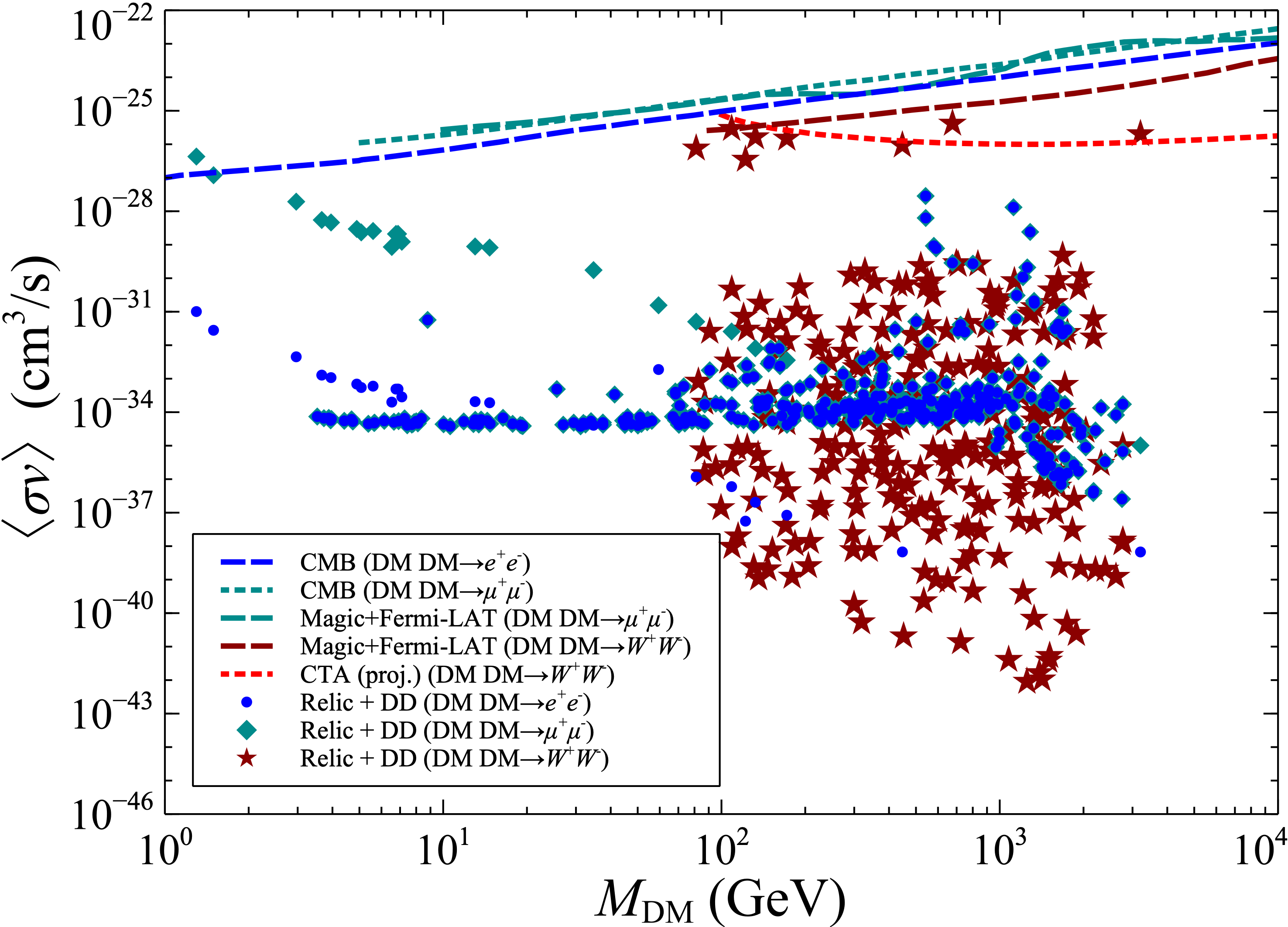}
    \caption{Case-I (\textit{left}), Case-II (\textit{right}): DM annihilation cross-section as a function of DM mass for ${\rm DM ~DM}\rightarrow e^+e^-$, ${\rm DM ~DM}\rightarrow\mu^+\mu^-$ and ${\rm DM ~DM}\rightarrow W^+W^-$ channels. The combined Magic and Fermi-LAT limits \cite{MAGIC:2016xys} are shown with dark cyan and dark red colored dashed lines for $\mu^+\mu^-$ and $W^+W^-$ channels, respectively. The projected sensitivity of CTA \cite{CTA:2020qlo} is shown with the red dotted line for the $W^+W^-$ channel. CMB constraints \cite{Slatyer:2015jla} are shown with blue dashed and dark cyan dashed lines for $e^+e^-$ and $\mu^+\mu^-$ channels, respectively.}\label{fig:idplot}
\end{figure*}

Beyond direct detection experiments, DM within the WIMP paradigm can also be explored through indirect detection experiments. These searches focus on detecting SM particles produced via DM annihilation or decay. Among the potential final-state particles, photons and neutrinos are particularly significant, as their neutrality and stability enable them to travel unimpeded from the source to the detector.  

Gamma rays, originating from electromagnetically charged final states, fall within the typical energy range expected for WIMP DM and can be observed using space-based telescopes such as the Fermi Large Area Telescope (Fermi-LAT) or ground-based observatories like MAGIC and HESS. By measuring the gamma-ray flux and incorporating standard astrophysical inputs, one can infer possible DM annihilation channels into various SM final states, such as $\mu^+\mu^-$, $\tau^+\tau^-$, and $W^+W^-$. Since DM particles do not directly couple to photons, the observed gamma rays arise from the decay of these charged particles. Constraints on DM properties can thus be derived by analyzing Fermi-LAT and MAGIC observations of dwarf spheroidal galaxies (dSphs)~\cite{MAGIC:2016xys}.  

Given that DM can annihilate into multiple final states, Fermi-LAT constraints on individual annihilation channels tend to be relatively weak in many scenarios. In Fig.~\ref{fig:idplot}, we present the DM annihilation cross-section into $\mu^+\mu^-$ and $W^+W^-$ channel as a function of DM mass along with the most stringent constraints from MAGIC+Fermi-LAT. {We also show the projected sensitivity of CTA \cite{CTA:2020qlo} for the $W^+W^-$ channel with red dotted line.} The parameter points shown in this figure correspond to those that satisfy relic density as well as direct detection constraints in both Case I and Case II. Clearly, all points in Fig.~\ref{fig:idplot}, which are consistent with relic density requirements as well as direct detection bounds, also remain within the limits set by indirect detection experiments. 

In addition to indirect search constraints, it is also important to consider the bounds arising from CMB measurements, which are sensitive to energy injection from DM annihilation around the time of recombination. These constraints are usually expressed through the quantity
$$
p_{\rm ann} = f_{\rm eff}(z)\,\frac{\langle\sigma v\rangle}{M_{\rm DM}}$$
where $f_{\rm eff}(z)$ denotes the redshift-dependent efficiency factor corresponding to a given annihilation channel. The Planck analysis places the upper bound 
$$p_{\rm ann} \lesssim 4.1\times10^{-28}~{\rm cm^3\,s^{-1}\,GeV^{-1}}$$ at $95\%$ C.L.~\cite{Slatyer:2015jla}. In Fig.~\ref{fig:idplot}, we show the corresponding CMB limits for the $e^+e^-$ and $\mu^+\mu^-$ channels (blue and dark-cyan dashed lines). Clearly, all points satisfying the correct relic density and direct detection constraints have annihilation cross sections into $e^+e^-$ that lie well below the CMB exclusion band, and hence are consistent with Planck bounds.
We also note that at lower DM masses, where both the CMB and indirect detection limits are particularly stringent, often approaching or excluding the canonical thermal cross-section, our scenario still remains viable. 
This is because the relic density in this mass range is dominantly determined by the DM annihilation into $ h_2 h_2$ via a $t$-channel process, while the branching to other channels is small. As DM annihilation to $h_2 h_2$ is a $p$-wave-suppressed process, the thermally averaged annihilation cross-section consequently decreases with temperature, making the late-time annihilation rate negligibly small and allowing the model to evade both CMB and indirect detection constraints.

\subsection{Collider Search}
\noindent\underline{\bf Collider Constraints on $Z_{\rm BL}$:}\\

In addition to constraints from relic density and direct or indirect DM searches, the $B-L$ gauge sector is subject to stringent experimental limits from collider experiments such as ATLAS, CMS, and LEP-II. Data from LEP-II~\cite{Carena:2004xs,Cacciapaglia:2006pk} impose a lower bound on the ratio of the new gauge boson mass to its coupling, requiring it to be at least 7 TeV. However, current LHC experiments have already surpassed the bounds set by LEP-II.  
In particular, searches for high-mass di-lepton resonances at ATLAS~\cite{ATLAS:2019erb} and CMS~\cite{CMS:2021ctt} have placed even tighter constraints on the additional gauge boson, significantly limiting the viable parameter space for the $B-L$ extension.

\begin{figure}[h]
	\centering	\includegraphics[scale=0.32]{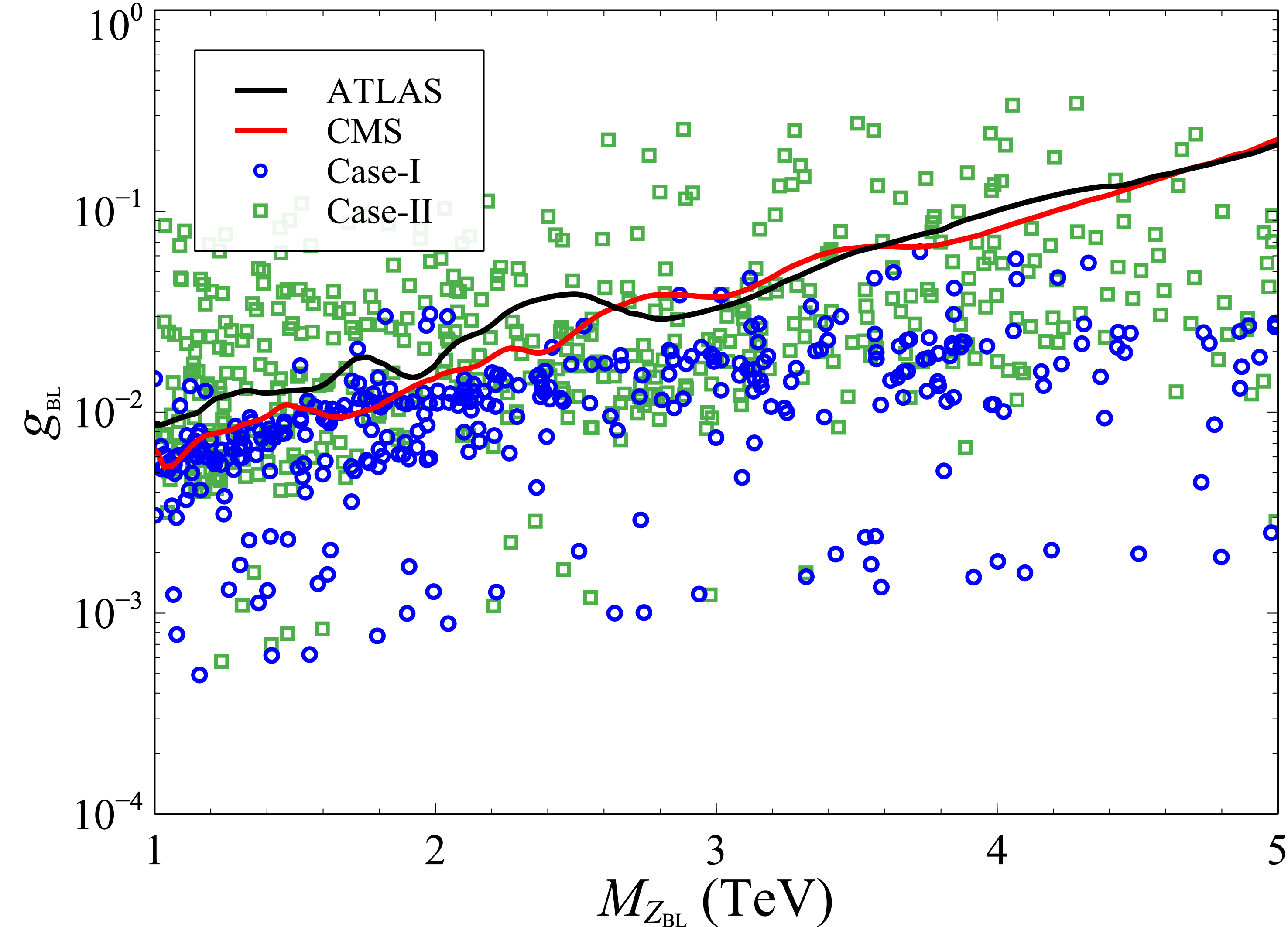}\caption{{The points satisfying relic density, direct detection, and indirect detection constraints are shown in the $g_{\rm BL}-M_{Z_{\rm BL}}$ plane for the case-I (case-II) with blue circles (green boxes). The constraints from ATLAS\cite{ATLAS:2019erb} and CMS \cite{CMS:2021ctt} are shown with black and red solid lines, respectively.}}\label{fig:gblvsmzblsum}
\end{figure}

To incorporate these collider constraints into our analysis, we follow the approach outlined in~\cite{Das:2021esm,Ghosh:2021khk}, where the maximum allowed value of the gauge coupling $g_{\rm BL}$ for a given gauge boson mass $M_{Z_{\rm BL}}$ is determined as  
\begin{eqnarray}
g^{\rm U.L.}_{\rm BL}=\sqrt{\frac{\sigma_{\rm Exp}}{\sigma_{\rm Th}/g^2_{\rm Th}}},
\end{eqnarray}
where $\sigma_{\rm Exp}$ represents the experimental upper limit on the production cross-section of the process $pp\rightarrow Z_{\rm BL}+X\rightarrow l^+l^-+X$, with $l = e, \mu$. The term $\sigma_{\rm Th}$ denotes the theoretical prediction of the same process within the model, computed for a benchmark gauge coupling $g_{\rm Th}$.  
It is important to note that, in our model, $Z_{\rm BL}$ has additional decay channels compared to the conventional $\rm B-L$ scenario. Consequently, the branching ratio of $Z_{\rm BL}$ into charged leptons ($l^+l^-$) is reduced, leading to weaker constraints on the $g_{\rm BL}-M_{Z_{\rm BL}}$ parameter space.

Finally, we present the parameter points that satisfy the relic density, direct detection, and indirect detection constraints in the $g_{\rm BL}-M_{Z_{\rm BL}}$ plane in Fig.~\ref{fig:gblvsmzblsum}, allowing us to examine the viable parameter space against collider constraints from ATLAS and CMS. The exclusion limits from CMS and ATLAS, derived using the method outlined above, are shown as red and black solid lines, respectively.  

Only parameter points that lie below both the red and black solid lines remain consistent with all relevant constraints (i.e., relic density, direct detection, indirect detection, and collider bounds from ATLAS and CMS). As evident from the figure, some of the previously allowed points are excluded once these collider constraints are imposed.

\vspace{0.2cm}
\noindent\underline{\bf Collider Signature of Triplet Fermion:}\\

\begin{figure}[h]
	\centering
\includegraphics[scale=0.35]{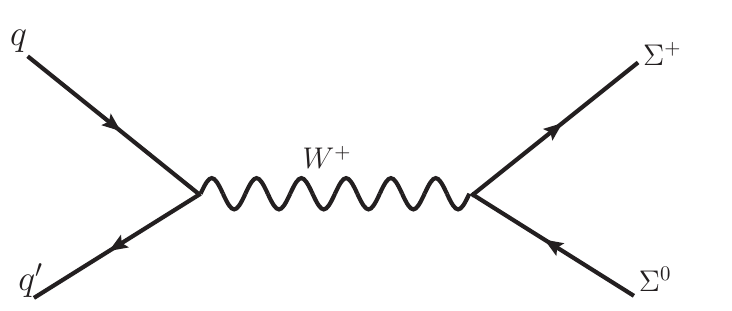}\\\includegraphics[scale=0.35]{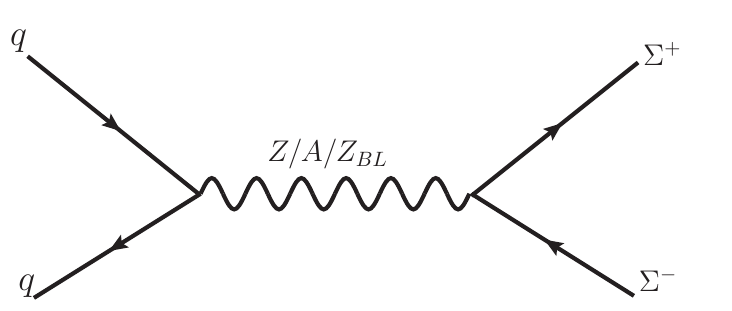}\includegraphics[scale=0.35]{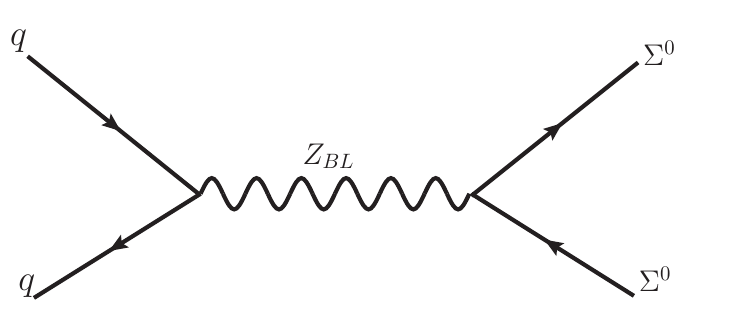}\caption{Feynmann diagrams for $\Sigma^{\pm,0}$ production through $p-p$ collision.}\label{fig:sigmaprod}
\end{figure}

\begin{figure}[h]
	\centering \includegraphics[scale=0.32]{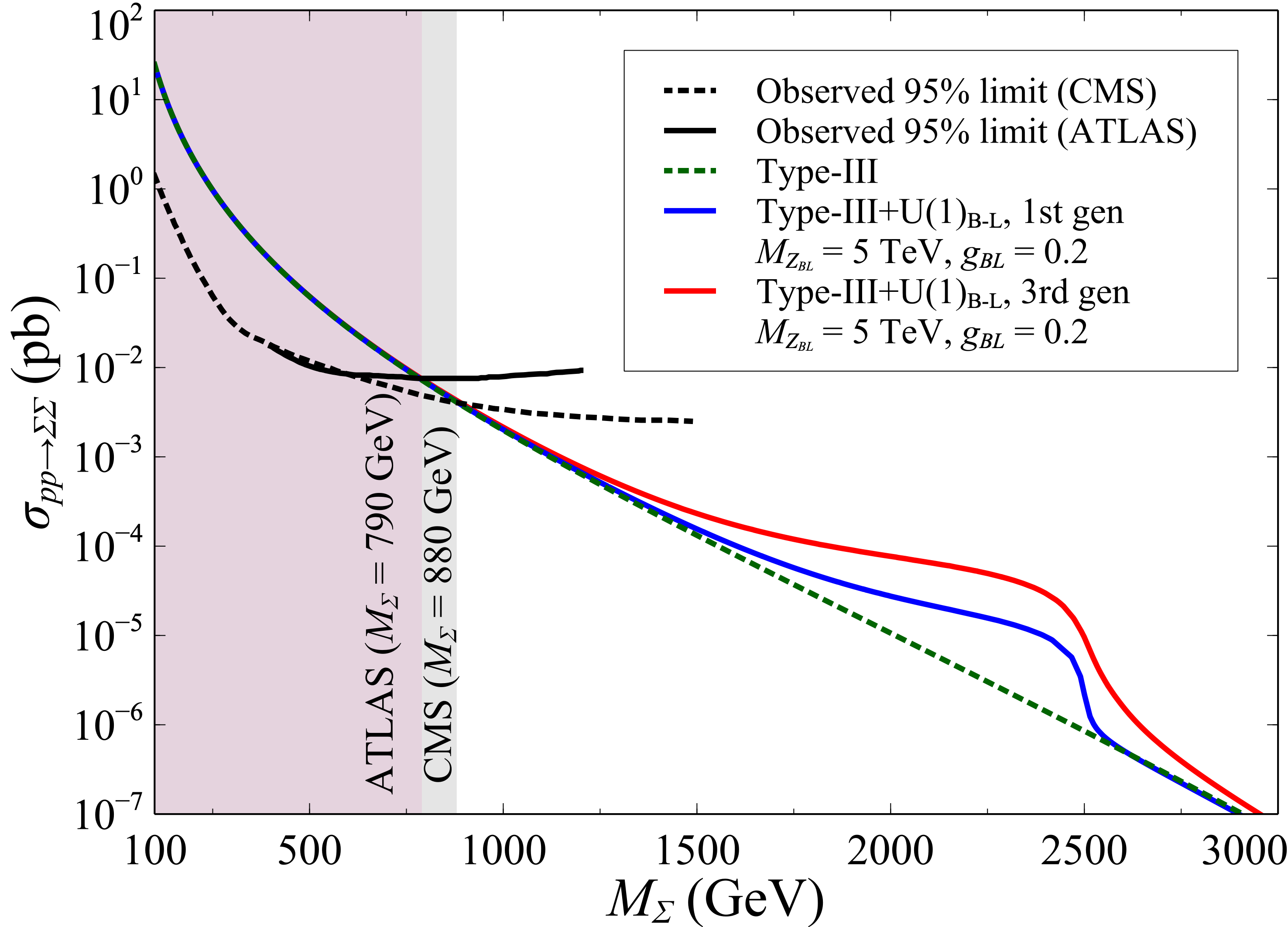}
	\caption{{Pair production of $\Sigma^{\pm, 0}$ at the 13 TeV LHC. The black dotted (solid) line represents the 95\% CL upper limit from the CMS~\cite{CMS:2019lwf} (ATLAS~\cite{ATLAS:2020wop}) experiment. The green dotted line shows the production cross-section in the absence of $Z_{\rm BL}$. The blue (red) line represents the production cross-section for $M_{Z_{\rm BL}}=5$ TeV and $g_{\rm BL}=0.2$, corresponding to the first (third) generation of triplet fermions.
}}\label{fig:sigmaCMS}
\end{figure}

In this section, we briefly discuss the detection prospects of this scenario at collider experiments. Fig.~\ref{fig:sigmaprod} illustrates the production channels of the fermion triplet in proton-proton collisions.
We compute the production cross-section of the triplet fermion in proton-proton collisions at a center-of-mass energy of $\sqrt{s} = 13$ TeV, shown as a green dotted line in Fig.~\ref{fig:sigmaCMS} as a function of $M_{\Sigma}$. As expected, the production cross-section decreases with increasing $M_{\Sigma}$. Existing collider searches at ATLAS~\cite{ATLAS:2020wop} and CMS~\cite{CMS:2019lwf} have already placed constraints on the triplet fermion mass, requiring it to be greater than 790 GeV and 880 GeV, as shown in Fig.~\ref{fig:sigmaCMS}.

In the conventional type-III seesaw scenario, the production of $\Sigma$ occurs primarily via SM gauge boson-mediated Drell-Yan processes. However, in a gauged $\rm B-L$ framework, the additional gauge boson $Z_{\rm BL}$ can significantly impact this pair production cross-section. This effect is illustrated in Fig.~\ref{fig:sigmaCMS} for benchmark values of $M_{Z_{\rm BL}} = 5$ TeV and $g_{\rm BL} = 0.2$ consistent with the ATLAS and CMS constraints on additional gauge boson mass and gauge coupling. We consider the production of the first generation of triplets as well as the third generation with the same mass. The third generation of triplets has a B-L charge twice that of the first generation, resulting in a larger cross-section than the latter. Our results show that the presence of $Z_{\rm BL}$ enhances the production cross-section, particularly for $M_{\Sigma} \gtrsim 1.1$ TeV.

At $M_{\Sigma} = 1.5$ TeV, the production cross-section in the presence of $Z_{\rm BL}$ is  
$
\sigma_{pp\rightarrow\Sigma\Sigma} = 1.56\times10^{-4}~( {2.31\times10^{-4}} )~\text{pb},
$
compared to  
$
\sigma_{pp\rightarrow\Sigma\Sigma} = 1.32\times10^{-4}~\text{pb}
$
in the pure type-III seesaw scenario. This corresponds to an enhancement of approximately $18\%$ $( {75\%} )$ for the first (third) generation of triplet fermions.  

Furthermore, as $M_{\Sigma}$ approaches $\approx 2.4$ TeV, where the resonance condition is satisfied for $M_{Z_{\rm BL}} = 5$ TeV, the enhancement becomes even more pronounced. In this case, the cross-section in the presence of $Z_{\rm BL}$ is  
$
\sigma_{pp\rightarrow\Sigma\Sigma} = 9.74\times10^{-6}~( {2.94\times10^{-5}} )~\text{pb},
$
whereas in its absence, it is  
$
\sigma_{pp\rightarrow\Sigma\Sigma} = 1.42\times10^{-6}~\text{pb}.
$
This results in a substantial enhancement of 586\% $( {1970\%} )$ for the first (third) generation.  

\begin{figure}[h]
	\centering \includegraphics[scale=0.32]{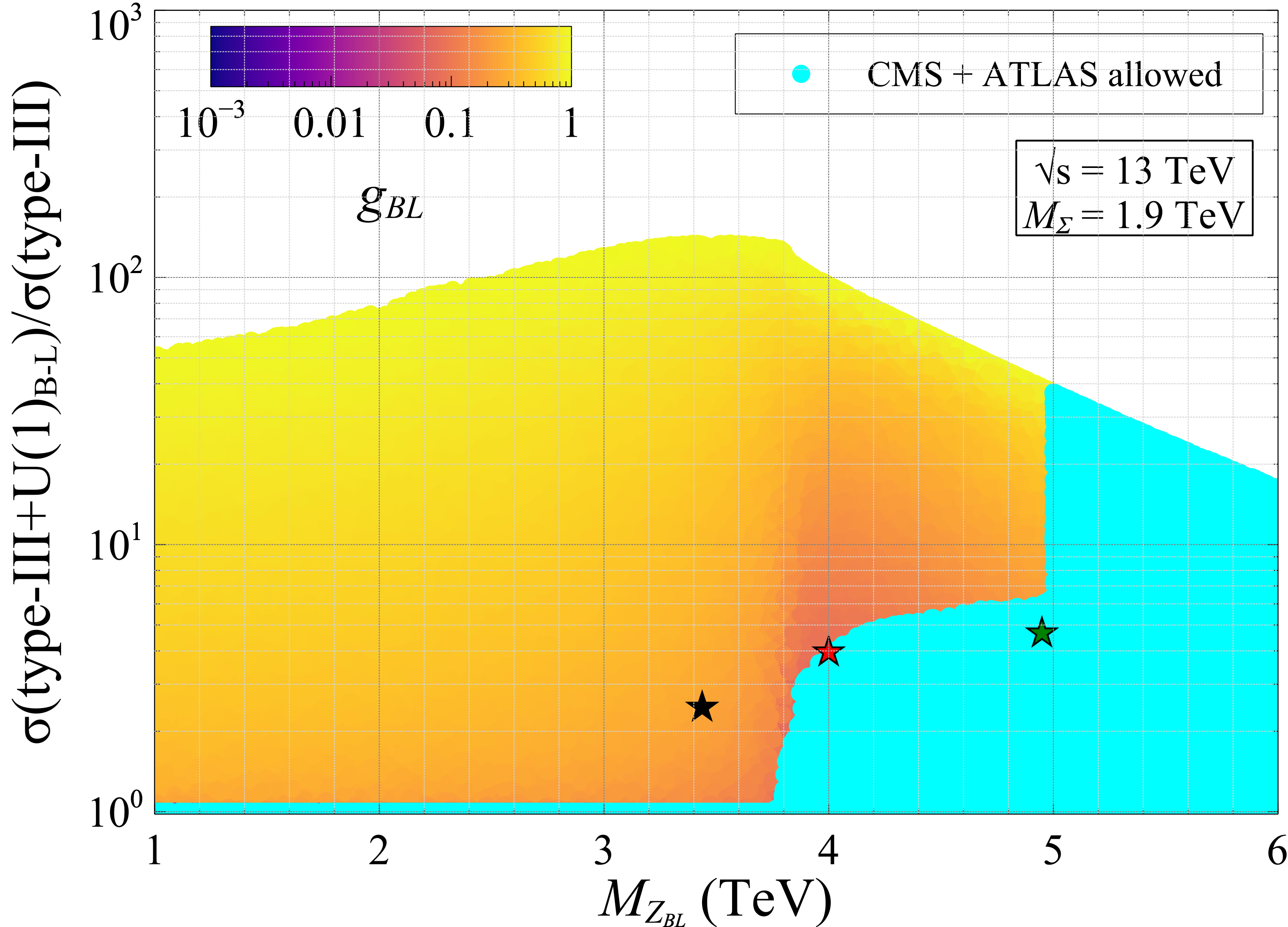}
	\caption{A comparison of $\Sigma^{\pm, 0}$ pair production cross-sections at the 13 TeV LHC in the minimal type-III seesaw model and its $U(1)_{\rm B-L}$-extended counterpart, focusing on third-generation couplings with $M_{\Sigma} = 1.9$ TeV. Regions of parameter space consistent with ATLAS~\cite{ATLAS:2019erb} and CMS~\cite{CMS:2021ctt} constraints are indicated by cyan data points. The stars correspond to three benchmarks as mentioned in Table \ref{tab:tab2}.}\label{fig:sigmaratCMS}
\end{figure}

To further examine the impact of $Z_{\rm BL}$ on the $\Sigma$ production cross-section at colliders, we conduct an extensive scan over $M_{Z_{\rm BL}}$ and $g_{\rm BL}$, computing the production cross-section at the LHC with $\sqrt{s} = 13$ TeV for a fixed $M_{\Sigma} = 1.9$ TeV for the third generation.  

In Fig.~\ref{fig:sigmaratCMS}, we present the relative enhancement in the production cross-section, defined as $$
{\sigma(\rm type\text{-}III + U(1)_{\rm B-L})}/{\sigma(\rm type\text{-}III)}
$$
as a function of $M_{Z_{\rm BL}}$, with the gauge coupling $g_{\rm BL}$ randomly varied within the range $[10^{-3},1]$. The figure illustrates that, away from the resonance, the production cross-section can increase by a factor of approximately 50, while near resonance, this enhancement can reach up to 140 times, as indicated by the gradient-colored points.

However, when we impose the ATLAS+CMS constraints on the $g_{\rm BL}-M_{Z_{\rm BL}}$ parameter space, the results are shown by the cyan points. Even after applying these constraints, we still observe an enhancement of up to 40 times. These results highlight the crucial role of the $Z_{\rm BL}$ gauge boson in modifying the collider phenomenology of type-III seesaw models, offering a potential signature for probing this scenario at future high-energy experiments. 

\begin{figure}[h]
    \centering
    \includegraphics[scale=0.41]{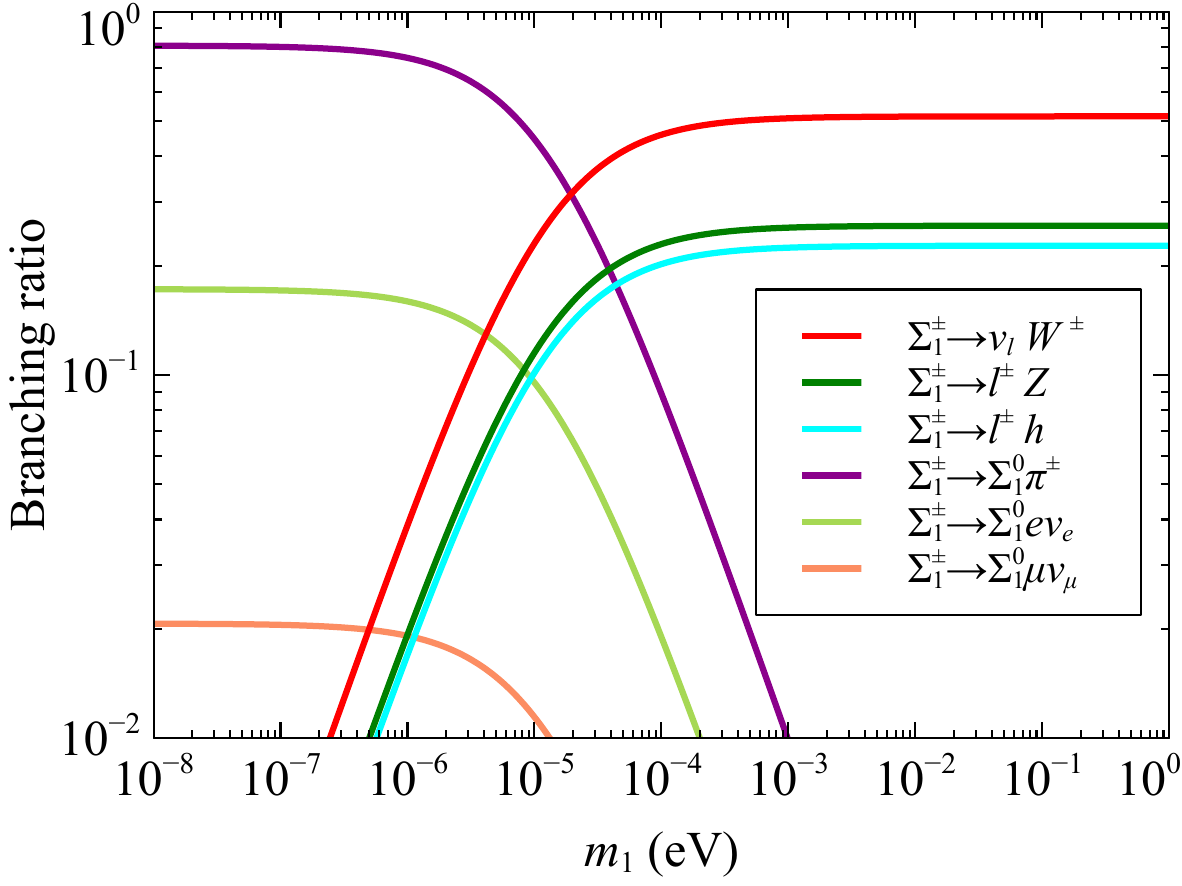}
    \caption{{Branching ratio of different decay modes of $\Sigma_1^\pm$ as a function of lightest neutrino mass in NH, summing over all lepton flavor final states. We fixed the triplet mass at 500 GeV.}}
    \label{fig:branching}
\end{figure}

We note that the exclusion limits shown in Fig.~\ref{fig:sigmaCMS} arise under the assumption that the heavy triplet fermions decay promptly into SM leptons and gauge/Higgs bosons.  Such prompt decays require sufficiently large mixing between the triplet and SM leptons, which ensures a decay length much shorter than the detector scale.  By contrast, in the regime where the mixing is extremely small, as considered in the following discussion, the charged component \(\Sigma^\pm_1\) becomes long-lived, and the prompt-decay bounds no longer apply.  Instead, one must rely on long-lived particle searches (e.g.\ disappearing charged tracks or displaced vertices). 

When the $\Sigma$-leptons mixing is small, once produced, the charged components $\Sigma_1^{\pm}$ predominantly decay via $\Sigma_1^0\pi^\pm$ arising from electroweak radiative corrections that induce a small mass splitting of $\Delta M\simeq 166$ MeV between the charged and neutral components of the triplet \cite{Cirelli:2005uq}.
This mass splitting enables the two-body pion mode, which accounts for nearly $98\%$ of the total decay width, while the subdominant three-body leptonic channels 
$\Sigma_1^{\pm} \rightarrow \Sigma_1^0 e^{\pm} \bar{\nu}_{e} (\nu_e)$ and 
$\Sigma_1^{\pm} \rightarrow \Sigma_1^0 \mu^{\pm} \bar{\nu}_{\mu} (\nu_\mu)$ 
contribute only $\sim 2\%$.
Additionally, due to the mixing of $\Sigma_1^\pm$ with the SM charged leptons, it can decay through the two-body processes such as: $\nu_lW^\pm,l^\pm Z,l^\pm h,$. In general, the mixing between the triplet fermions $\Sigma$ and the SM charged leptons is controlled by the structure of the neutrino Yukawa matrix, which in turn depends on the choice of the orthogonal matrix $\mathcal{R}$ in the Casas–Ibarra parametrization \cite{Casas:2001sr}. For the special choice $\mathcal{R} = \mathcal{I}_{3\times3}$, the branching ratios of the two-body decays $\Sigma_1^\pm \;\to\; \nu_\ell W^\pm,\; \ell^\pm Z,\; \ell^\pm h$, are sensitive to the lightest neutrino mass $m_1$ (assuming normal hierarchy). On the other hand, the decay channels mediated purely by electroweak corrections, such as $\Sigma_1^{\pm} \rightarrow \Sigma_1^0 \pi^{\pm}$ and the leptonic three-body decays mentioned above, remain essentially independent of $m_1$\footnote{A detailed discussion of the decay widths can be found in \ref{app:tripletdecay}.}. 

In Fig.~\ref{fig:branching}, we present the branching ratios of $\Sigma_1^{\pm}$ to different final states for $M_{\Sigma_1} = 500~\text{GeV}$, assuming a degenerate triplet mass spectrum and $\mathcal{R} = \mathcal{I}$. 
The best-fit values of neutrino oscillation parameters used in this analysis are listed in Table~\ref{tab:tab3}. 
We observe that for larger $m_1$ values, the two-body decays $\nu W^{\pm}$, $\ell^{\pm} Z$, and $\ell^{\pm} h$ dominate, while for smaller $m_1 \lesssim \mathcal{O}(2\times10^{-5})~\text{eV}$, the pion channel $\Sigma_1^{\pm} \to \Sigma_1^0 \pi^{\pm}$ becomes dominant. 
In the latter case, $\Sigma_1^{\pm}$ behaves as a long-lived charged particle, potentially giving rise to disappearing charged track signatures at colliders.
\begin{figure}[h]
	\centering
    \includegraphics[scale=0.41]{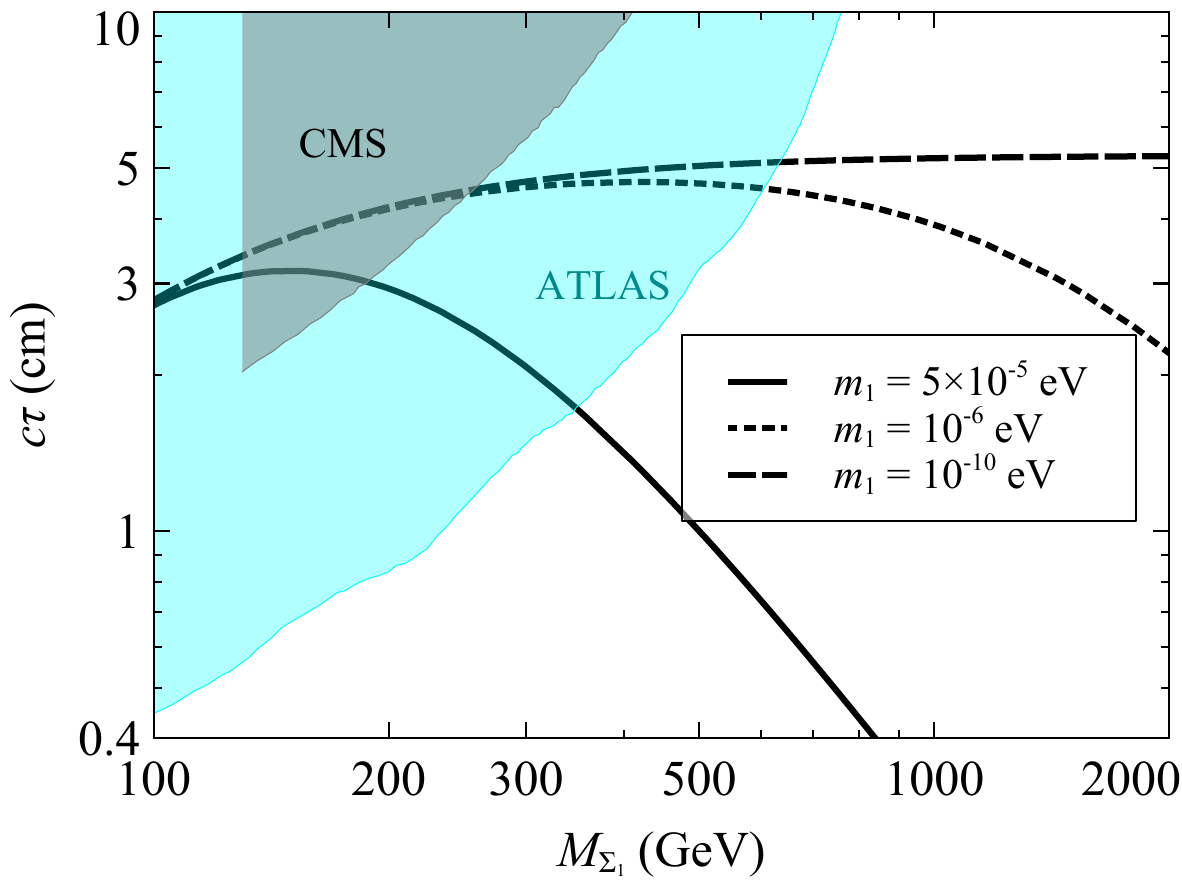}
	\caption{{The total decay length of $\Sigma_1^{\pm}$ versus its mass, compared with the ATLAS~\cite{ATLAS:2022rme} and CMS~\cite{CMS:2018rea} bounds on disappearing charged track searches at a center-of-mass energy of 13 TeV.}}\label{ctauvsmass}
\end{figure}
The ATLAS~\cite{ATLAS:2022rme} and CMS~\cite{CMS:2018rea} searches for disappearing charged tracks have imposed stringent limits on long-lived charged particles such as $\Sigma_1^{\pm}$. For instance, current data exclude $\Sigma_1^{\pm}$ masses below $640~\text{GeV}$ for $m_1=10^{-10}~\text{eV}$, and $606~\text{GeV}$ for $m_1=10^{-6}~\text{eV}$.  In Fig.~\ref{ctauvsmass}, we present the decay length of $\Sigma_1^\pm$ for three choices of $m_1:\{5\times10^{-5},10^{-6},10^{-10}\}$eV as a function of $\Sigma_1^\pm$ mass. 
Given that the electroweak-induced mass splitting $\Delta M \simeq 166~\text{MeV}$ naturally leads to a macroscopic decay length, such disappearing track signatures serve as a promising avenue to test this model at present and future collider experiments.
\begin{figure}[h]
\centering \includegraphics[scale=0.41]{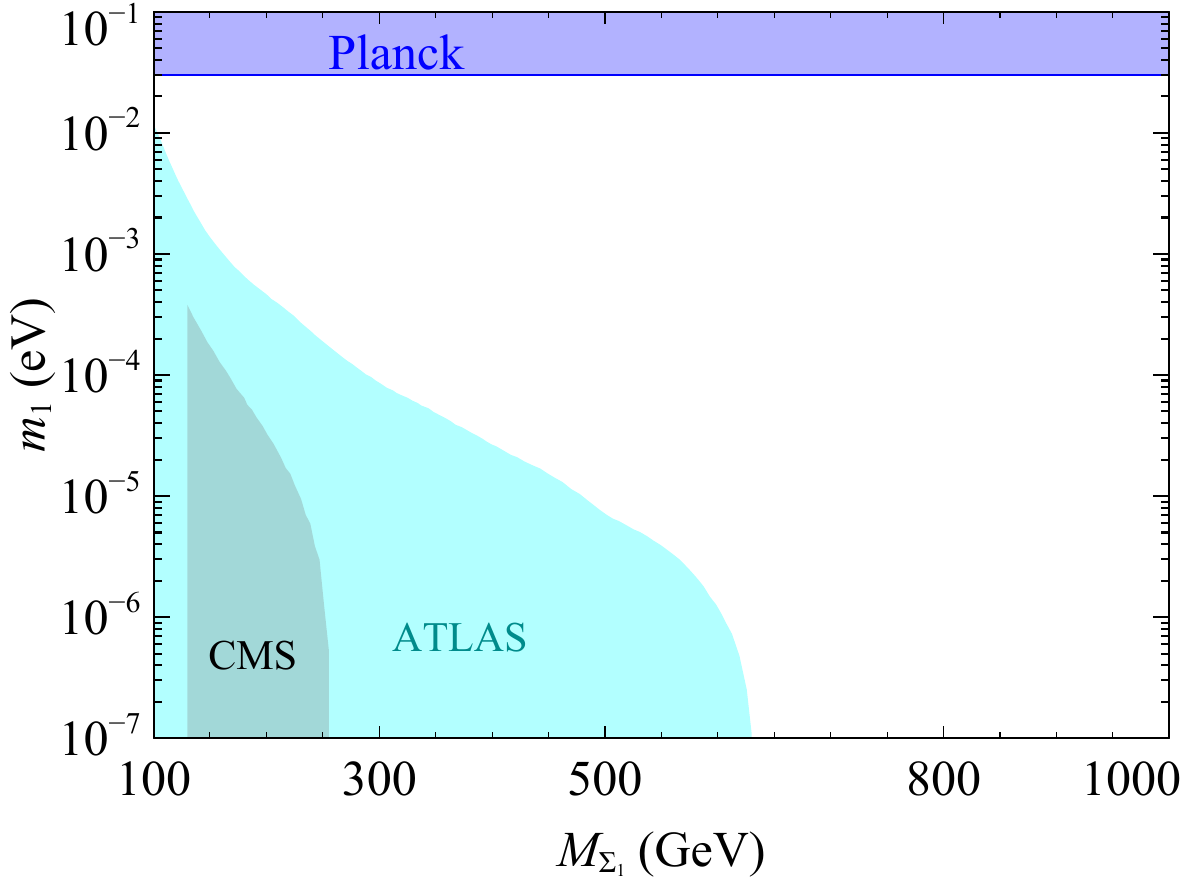}
\caption{{Allowed parameter space (white region) from collider experiments in the plane of $m_1$ vs $M_{\Sigma_1}$. Neutrino oscillation data are fixed at the best fit values as mentioned in Table \ref{tab:tab3}.}}
\label{fig:m1vsmsigma}
\end{figure}
In Fig. \ref{fig:m1vsmsigma}, we show the exclusion region from ATLAS and CMS experiments in the plane of $m_1$ vs $M_{\Sigma_1}$. The blue region represents the upper limit on the mass of the lightest SM neutrino from Planck data. Here, we have used the best-fit values of the neutrino oscillation data.

We emphasize that the flavor structure of the heavy–light mixing in the type-III seesaw depends crucially on the choice of the orthogonal matrix \(\mathcal{R}\) in the Casas–Ibarra parametrization.  While in the earlier part of our analysis we set \(\mathcal{R} = \mathcal{I}_{3\times3}\) for definiteness, this choice corresponds to a particular alignment of the Yukawa couplings and in general need not hold.  Relaxing this assumption changes the mixing strength between the triplet fermions and the SM leptons, which in turn will affect the branching ratios and lifetimes of \(\Sigma^\pm\).  Hence conclusions based on \(\mathcal{R} = \mathcal{I}\) should be interpreted as representative, not universal.  

\begin{figure}[h]
	\centering
    \includegraphics[scale=0.41]{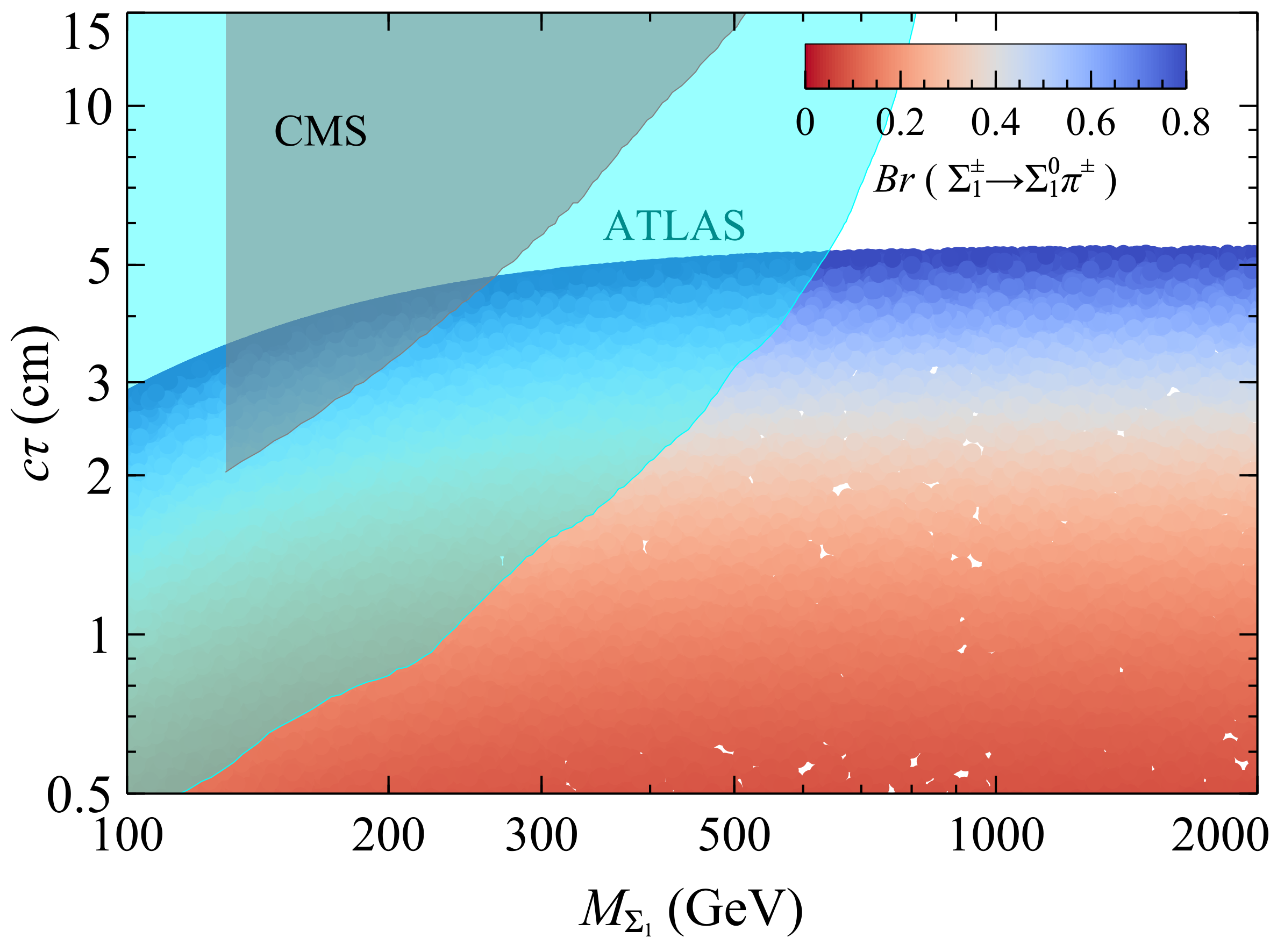}
	\caption{The total decay length of $\Sigma_1^{\pm}$ versus its mass, compared with the ATLAS~\cite{ATLAS:2022rme} and CMS~\cite{CMS:2018rea} bounds on disappearing charged track searches at a center-of-mass energy of 13 TeV considering a general rotation matrix.}\label{fig:ctauvsmass3}
\end{figure}

In order to check the robustness of our conclusions, we have also performed a numerical scan over a general orthogonal matrix
$\mathcal{R} = \mathcal{O}_{23}\,\mathcal{O}_{13}\,\mathcal{O}_{12}$, with complex rotation angles $\theta_{ij}=x_{ij}+iy_{ij}$, varying over a broad range $x_{ij},y_{ij}\in [10^{-4},1]$. Also $m_1$ is varied in the range [$10^{-10},1$] eV and $M_{\Sigma_1}$ in the range $[100,2000]$ GeV.  For each choice of $\mathcal{R}$, we recomputed the decay branching ratio of $\Sigma_1^\pm \rightarrow \Sigma_1^0 \, \pi^\pm$ as well as the total decay length. We find that the result is no longer sensitive to $m_1$. The result is summarized in Fig.~\ref{fig:ctauvsmass3}, where the color code indicates the branching fraction into the pion channel, and demonstrates that, across a wide span of $\mathcal{R}$, the qualitative conclusion remains valid that in regions where the mixing is small, the pion channel dominates and $\Sigma^\pm$ becomes long-lived, yielding disappearing-track signatures at colliders.  

\vspace{0.5cm}
\noindent\underline{\textit{Constraints from neutrinoless double beta decay:}}\\
Neutrinoless double beta ($0\nu\beta\beta$) decay provides one of the most sensitive probes of the Majorana nature of neutrinos and offers complementary information about the absolute neutrino mass scale. The observable relevant to $0\nu\beta\beta$ decay experiments is the effective Majorana mass, defined as
\begin{eqnarray}
    \label{eq:mbb}m_{\beta\beta}&=&m_1\cos^2\theta_{12}\cos^2\theta_{13}e^{2i\varphi_1}+m_2\sin^2\theta_{12}\cos^2\theta_{13}e^{2i\varphi_2}\nonumber\\&&+m_3\sin^2\theta_{13} ,
\end{eqnarray} 
where $\varphi_{1,2}$ denote the two Majorana $CP$ phases that appear in the PMNS matrix. In Eq.~(\ref{eq:mbb}), the only free parameters are $\{m_{\rm lightest}, \varphi_1, \varphi_2\}$, where $m_{\rm lightest}=m_1$ for the normal hierarchy and $m_{\rm lightest}=m_3$ for the inverted hierarchy. 
For the two mass orderings, the neutrino masses can be expressed as
\begin{align*}
	&\text{NH: } 
	m_2 = \sqrt{\Delta m_{21}^2 + m_1^2}, \quad 
	m_3 = \sqrt{\Delta m_{31}^2 + m_1^2}, \nonumber\\
	&\text{IH: } 
	m_1 = \sqrt{\Delta m_{23}^2 - \Delta m_{21}^2 + m_3^2}, \quad 
	m_2 = \sqrt{\Delta m_{23}^2 + m_3^2}.
\end{align*}
We fix the neutrino mixing angles and mass-squared differences at their best-fit values given in Table~\ref{tab:tab3}, and vary the Majorana phases $\varphi_{1,2}\in[0,\pi]$ and the lightest neutrino mass $m_{\rm lightest}\in[10^{-10},1]~{\rm eV}$ to obtain the allowed range of $m_{\beta\beta}$.
\begin{table}[h]
		\centering
			\begin{tabular}{|c|c|c|}
				\hline Parameters & Best fit & $3\sigma$ range\\
				\hline
				$\Delta m_{21}^2[10^{-5}\rm eV^2]$&7.5&6.94–8.14\\
                \hline
				$\Delta m_{31}^2[10^{-3}\rm eV^2](NH)$&2.55&2.47–2.63\\
                \hline
				$\Delta m_{23}^2[10^{-5}\rm eV^2](IH)$&2.45&2.37–2.53\\
                \hline
				$\sin^2\theta_{12}$&0.318&0.271–0.369\\
                \hline
				$\sin^2\theta_{23}\rm (NH)$&0.574&0.434–0.61\\
               \hline
                $\sin^2\theta_{23}\rm (IH)$&0.578&0.433–0.608\\
                \hline
				$\sin^2\theta_{13}\rm (NH)$&0.022&0.02–0.02405\\
                \hline
				$\sin^2\theta_{13}\rm (IH)$&0.02225&0.02018–0.02424\\
                \hline
				$\delta\rm (NH)$&194\degree&128\degree–359\degree\\
                \hline
				$\delta\rm (IH)$&284\degree&200\degree–353\degree\\
				\hline
		\end{tabular}
		\caption{The best fit values and the $3\sigma$ ranges of the neutrino oscillation parameters\cite{deSalas:2020pgw}.}
		\label{tab:tab3}
	\end{table}
\begin{figure}[h]
\centering \includegraphics[scale=0.41]{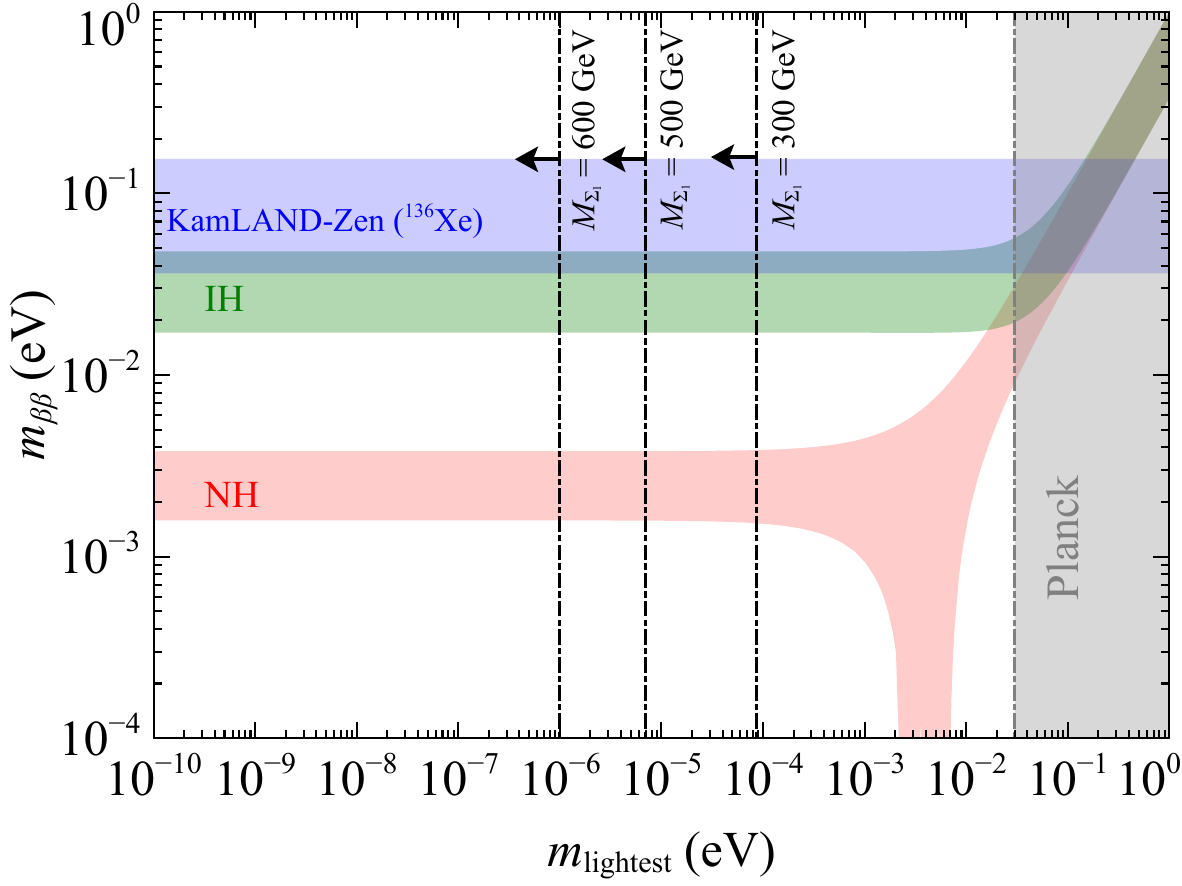}
\caption{{Effective neutrino mass as a function of lightest neutrino mass for NH (red band) and IH (green band) considering the best fit values of the neutrino oscillation parameters as mentioned in Table \ref{tab:tab3}. Vertical lines represent ATLAS exclusions (regions left to each line are excluded) for different $\Sigma_1$ masses.}}
\label{fig:meevsm1}
\end{figure}
The resulting variation of $m_{\beta\beta}$ as a function of $m_{\rm lightest}$ is shown in Fig.~\ref{fig:meevsm1}. The red (green) band corresponds to the NH (IH) spectrum, obtained by scanning over the allowed range of Majorana phases. The gray-shaded region represents the cosmological upper limit on the sum of neutrino masses from the Planck data~\cite{Planck:2018vyg}, while the blue-shaded area denotes the exclusion region from KamLAND-Zen \cite{KamLAND-Zen:2022tow}. The black dash-dotted vertical lines represent the ATLAS constraints, with the regions to their left excluded for the corresponding values of the heavy fermion mass $M_{\Sigma_1}$ in the NH case. We note that similar constraints also arise for the IH scenario, 
where the contributions are primarily governed by the triplet state $\Sigma_3$, and the allowed parameter space depends sensitively on its mass and mixing pattern.
From Fig.~\ref{fig:meevsm1}, we observe that the predicted $m_{\beta\beta}$ values may lie within the sensitivity reach of upcoming $0\nu\beta\beta$ experiments. The combined cosmological, collider, and neutrinoless double beta decay bounds thus provide a complementary set of constraints on the model parameter space.

\section{Stochastic Gravitational Waves from first-order phase transition}\label{sec:gw}

The model under consideration also presents intriguing cosmological detection prospects through stochastic gravitational waves, offering a complementary probe of its viability. These gravitational wave signatures originate from the phase transition associated with the spontaneous breaking of $U(1)_{\rm B-L}$. This symmetry is broken when the scalars $\Phi_{1,2,3},\Phi$ acquire vacuum expectation values.  

A first-order phase transition (FOPT) can occur if the true vacuum, where $U(1)_{\rm B-L}$ is broken, has a lower energy density than the high-temperature false vacuum, with a potential barrier separating them. Identifying the parameter space where an FOPT occurs requires analyzing the shape of the effective potential, including higher-order corrections along with the thermal corrections.  
In our scenario, four singlet scalars contribute to the breaking of $U(1)_{\rm B-L}$. To simplify the analysis, we consider an effective field, $\phi'$, which is a linear combination of these four fields, as the primary driver of the FOPT. A full treatment incorporating all four scalars simultaneously is beyond the scope of this work.

	\begin{table*}[t]
		\centering
		\resizebox{18cm}{!}{
			\begin{tabular}{
					|c|c|c|c|c|c|c|c|c|c|c|c|c|c|c|c|c|c|
				}
				\hline BPs&$M_{Z_{\rm BL}}\rm(GeV)$&$g_{_{\rm BL}}$ & $\lambda_{\phi^\prime}$ &$u(\rm TeV)$ & $v(\rm TeV)$ &$T_C (\rm TeV)$ & $T_n(\rm TeV)$ & $\alpha$ & $\beta/\mathcal{H}$ &$M_{\rm DM} (\rm GeV)$&$\sin_{\theta_{h\phi}}$ &$\rm Relic$&$\rm DD$ &$\rm Collider$&  $\Delta\sigma/\sigma\%$\\
				\hline
				BP1($\textcolor{green}{\star}$)&4949.29&0.18& 0.01 & 6 & 10.3925& 2.98 & 1.38414 & 0.150& 193.417&15&$10^{-5}$&\textcolor{blue}{\Checkmark}&\textcolor{blue}{\Checkmark}& \textcolor{blue}{\Checkmark}&366.278\%\\
                \hline
				BP2($\textcolor{red}{\star}$)& 4000&0.0814& 0.0005 & 10.721&18.5698 & 2.575& 0.86265 & 0.510 & 521.539&48.25&$10^{-5}$&\textcolor{blue}{\Checkmark}&\textcolor{blue}{\Checkmark}& \textcolor{blue}{\Checkmark}&295.904\%
                \\
                \hline
				BP3($\textcolor{black}{\star}$)& 3437.01&0.25& 0.002 & 3&5.19624 & 3.32& 3.15062 & 0.00955 & 4370.18&4.20&$10^{-5}$&\textcolor{blue}{\Checkmark}&\textcolor{blue}{\Checkmark}&\textcolor{red}{\XSolidBrush}& 145.851\% \\
				\hline
		\end{tabular}}
		\caption{Benchmark points satisfying DM relic, direct, and indirect detection constraints and giving observable GW signatures. The last column corresponds to the $\Sigma$ production cross-section enhancement due to the presence of $Z_{\rm BL}$ for $M_{\Sigma}=1.9$ TeV.}
		\label{tab:tab2}
	\end{table*}
    
\begin{figure*}[tbh]
	\centering	\includegraphics[scale=0.39]{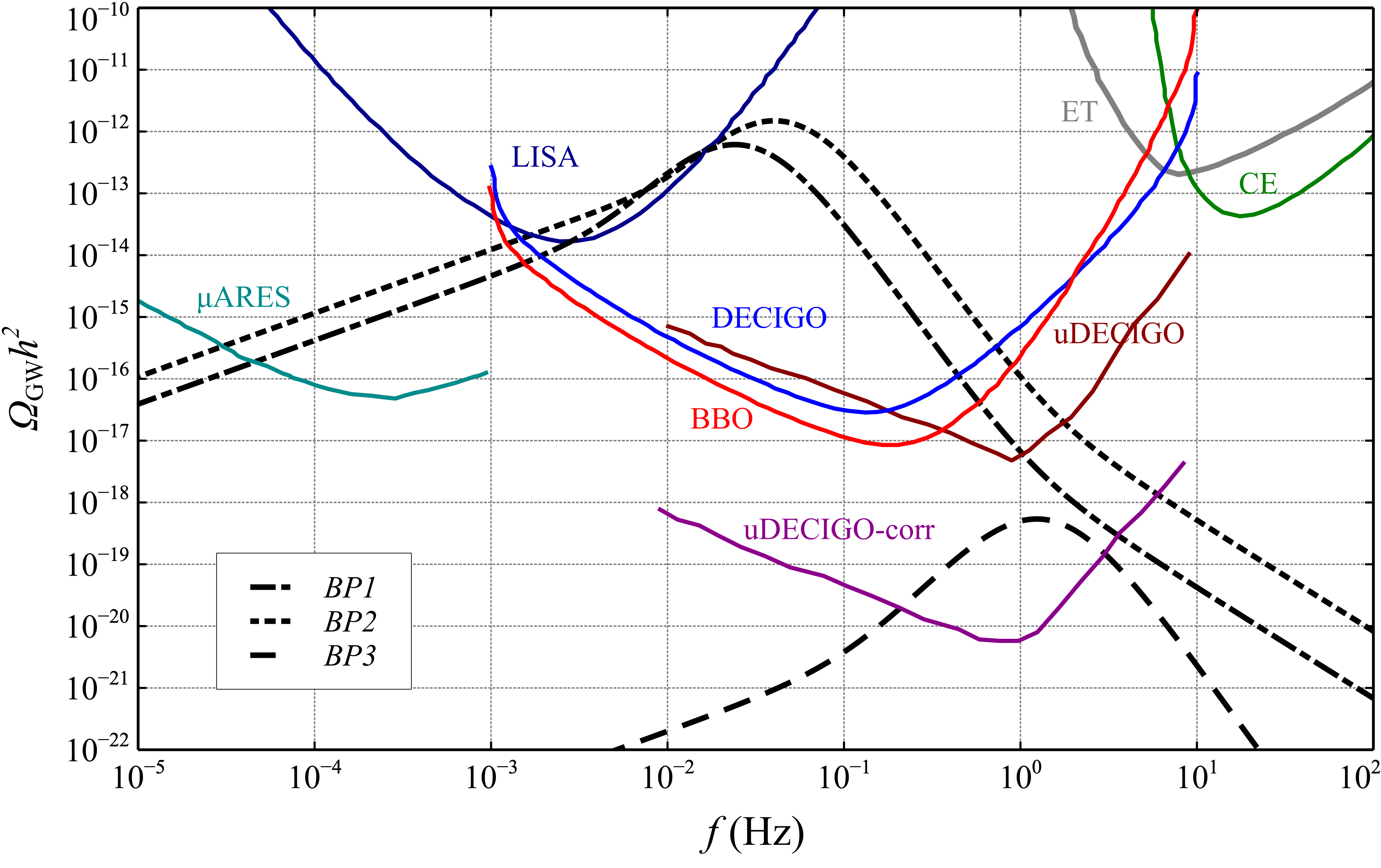}
	\caption{Gravitational wave spectrum from first-order phase transition for three benchmarks as shown in Table \ref{tab:tab2}. The sensitivities of different GW experiments are shown in solid-colored lines.}\label{fig:gwsprectrum}
\end{figure*}

As detailed in \ref{app:fopt}, we compute the complete effective potential, incorporating the tree-level potential $V_{\rm tree}$, the one-loop Coleman-Weinberg correction $V_{\rm CW}$ \cite{Coleman:1973jx}, and finite-temperature corrections~\cite{Dolan:1973qd,Quiros:1999jp}.  
The critical temperature $T_c$, at which the potential develops two degenerate minima $(0, v_c)$, is determined by analyzing the temperature evolution of the potential. The ratio $v_c/T_c$ serves as the order parameter, where a larger value indicates a stronger first-order phase transition (FOPT).  
The FOPT proceeds via quantum tunneling, with the tunneling rate estimated by calculating the bounce action $S_3$ following the approach in~\cite{Linde:1980tt, Adams:1993zs}. The nucleation temperature $T_n$ is then determined by equating the tunneling rate to the Hubble expansion rate of the universe, $\Gamma (T_n) = \mathcal{H}^4(T_n)$.  

We then compute the key parameters necessary to estimate the stochastic gravitational wave (GW) spectrum originating from bubble collisions~\cite{Turner:1990rc,Kosowsky:1991ua,Kosowsky:1992rz,Kosowsky:1992vn,Turner:1992tz}, sound waves in the plasma~\cite{Hindmarsh:2013xza,Giblin:2014qia,Hindmarsh:2015qta,Hindmarsh:2017gnf}, and plasma turbulence~\cite{Kamionkowski:1993fg,Kosowsky:2001xp,Caprini:2006jb,Gogoberidze:2007an,Caprini:2009yp,Niksa:2018ofa}.  
The two critical parameters for estimating the GW signal are the latent heat released relative to the radiation energy density $(\rho_{\rm rad})$ and the duration of the phase transition. These are expressed in terms of  $\alpha(T_n)$ and ${\beta}/{{\mathcal{H}}(T_n)}$ which determine the strength of the FOPT and are given by:  
\begin{eqnarray}
    \alpha(T_n)=\frac{\rho_{\rm vac}(T_n)}{\rho_{\rm rad}(T_n)},
\end{eqnarray}
where
\begin{eqnarray}
\rho_{\rm vac}(T)&=&V_{\rm eff}(\phi_{\rm false},T)-V_{\rm eff}(\phi_{\rm true},T)-\nonumber\\&&
T\frac{\partial}{\partial T}[V_{\rm eff}(\phi_{\rm false},T)-V_{\rm eff}(\phi_{\rm true},T)],
\end{eqnarray}
and
\begin{eqnarray}
\rho_{\rm rad}(T)=\frac{\pi^2}{30}g_{*}T^4,
\end{eqnarray}
where $g_*$ is the number of d.o.f when the bubbles are nucleated. The parameter $\beta$ is the inverse of the duration of PT, given as
\begin{eqnarray}
\frac{\beta}{\mathcal{H}(T_n)}\simeq T_n\frac{d}{dT}\bigg(\frac{S_3}{T}\bigg)\bigg|_{T=T_n}
\end{eqnarray}

As discussed earlier, the energy density of stochastic gravitational waves (GWs) receives contributions from three primary sources: bubble wall collisions, sound waves in the plasma, and magnetohydrodynamic (MHD) turbulence. The total GW energy spectrum can be approximated as the sum of these contributions:  
\begin{eqnarray}
\Omega_{\rm GW}h^2 \approx \Omega_{\rm col} h^2 + \Omega_{\rm sw} h^2 + \Omega_{\rm turb} h^2.
\end{eqnarray}
The details of these contributions are provided in \ref{app:gw}.  

We compute the GW spectrum generated by the first-order phase transition for three benchmark points that satisfy the correct relic density constraint as well as the constraints from DM direct and indirect detection experiments, as listed in Table~\ref{tab:tab2}. The corresponding GW spectra are shown in Fig.~\ref{fig:gwsprectrum}, along with the sensitivity curves of various upcoming GW experiments, such as BBO\cite{Yunes:2008tw}, CE\cite{LIGOScientific:2016wof}, DECIGO, U-DECIGO, U-DECIGO-corr~\cite{Kudoh:2005as,Adelberger:2005bt,Yagi:2011wg,Kawamura:2020pcg}, ET\cite{Punturo:2010zz}, LISA\cite{LISA:2017pwj}, $\mu$ARES\cite{Sesana:2019vho}, covering a broad frequency range.  

Our results indicate that all three benchmark points can be probed by future GW detectors. However, one of them (BP3) is already excluded by collider constraints on the $M_{Z_{\rm BL}}-g_{\rm BL}$ parameter space. Interestingly, all three benchmark points also predict a significantly enhanced production cross-section\footnote{In Table \ref{tab:tab2}, $\Delta\sigma/\sigma\equiv\frac{
\sigma(\rm type\text{-}III + U(1)_{\rm B-L})-\sigma(\rm type\text{-}III)}{\sigma(\rm type\text{-}III)}$, where $\sigma(\rm type\text{-}III)$ represents the production cross-section of $\Sigma$ in type-III seesaw, and $\sigma(\rm type\text{-}III + U(1)_{\rm B-L})$ represents the production cross-section of $\Sigma$ in an $U(1)_{\rm B-L}$ extended type-III seesaw model.} for the triplet fermion at colliders in the presence of the $Z_{\rm BL}$ boson. {The $BP1,BP2, BP3$ are shown with green, red and black colored stars in Fig. \ref{fig:sigmaratCMS}.} These findings highlight the complementary detection prospects of this scenario, demonstrating its rich and intriguing phenomenology across both cosmological and collider experiments.

\section{Conclusion}\label{sec:concl}

In this work, we have proposed a gauged $U(1)_{\rm B-L}$ extension of the Standard Model embedded within the type-III seesaw framework, addressing the dual challenges of neutrino mass generation and dark matter (DM) within a unified framework. By introducing $SU(2)_L$ triplet fermions to generate neutrino masses through the type-III seesaw mechanism and additional chiral fermions to cancel gauge anomalies, we establish a natural connection between neutrino physics and DM. The residual $Z_2$ symmetry arising from $U(1)_{\rm B-L}$ breaking stabilizes the Dirac fermion formed by the anomaly-canceling chiral fermions, providing a viable DM candidate.

A comprehensive analysis of the DM phenomenology reveals that the model successfully reproduces the observed relic density while evading constraints from direct detection experiments such as LZ, XENONnT, and indirect searches by Fermi-LAT and MAGIC. The interplay between the new scalar mediator and $Z_{\rm BL}$-mediated interactions governs the DM phenomenology, with the possibility of DM annihilation into the $\rm B-L$ scalar significantly expanding the viable parameter space, particularly for light DM masses below the SM Higgs resonance ($M_{\rm DM} \lesssim M_{h_2}/2$). Crucially, the presence of the $U(1)_{\rm B-L}$ gauge boson enhances the production cross-section of triplet fermions at colliders by roughly a factor of $10$, while disappearing charged track signatures from their decays provide a distinctive experimental probe of the model. We find that the decay pattern of the charged triplet is strongly correlated with the lightest neutrino mass, leading to potentially long-lived charged tracks at colliders. Current searches already place meaningful bounds on this parameter space, while a large region remains accessible to upcoming experimental facilities. The complementary sensitivity of neutrinoless double beta decay experiments further strengthens the overall testability of the framework.

The model also predicts a unique cosmological signatures through stochastic gravitational waves (GWs) generated during the first-order phase transition (FOPT) of $U(1)_{\rm B-L}$ symmetry breaking. Our analysis demonstrates that the resulting GW spectrum lies within the sensitivity ranges of upcoming observatories like LISA, DECIGO, and the Einstein Telescope, establishing GWs as a complementary probe of the model’s symmetry-breaking dynamics. The correlation between the collider signature {\it i.e.} enhanced triplet production with the GW parameters ($\alpha$ and $\beta$) and thus the predicted GW spectrum underscores the multi-faceted testability of this framework.

In summary, this gauged $U(1)_{\rm B-L}$ type-III seesaw model achieves a compelling synthesis of neutrino mass generation, rich DM phenomenology, and interesting collider and cosmological detection prospects. 
Future results from high-luminosity colliders, next-generation DM detectors, and GW observatories will critically assess its viability, potentially illuminating new physics beyond the Standard Model.

\begin{acknowledgements}
S.M. acknowledges the financial support from National Research Foundation(NRF)
grant funded by the Korea government (MEST) Grant No. NRF-2022R1A2C1005050. P.K.P. would like to acknowledge the Ministry of Education, Government of India, for providing financial support for his research via the Prime Minister’s Research Fellowship (PMRF) scheme. The works of N.S. and P.S. are supported by the Department of Atomic
Energy-Board of Research in Nuclear Sciences, Government of India (Grant No. 58/14/15/2021- BRNS/37220). S.M. and P.K.P. would like to thank Arnab Chaudhuri and Indrajit Saha for useful discussions on FOPT and gravitational waves.	
\end{acknowledgements}

\appendix

{\section{Triplet fermion decay modes}\label{app:tripletdecay}
The electroweak correction induces a small mass splitting between the charged and neutral components of the triplet. This mass splitting is given by~\cite{Cirelli:2005uq}:  
\begin{equation}
	\Delta M= \frac{\alpha_2 M_{\Sigma}}{4\pi}\left[s_w^2 f\left(\frac{M_Z}{M_\Sigma}\right) + f\left(\frac{M_W}{M_\Sigma}\right) - f\left(\frac{M_Z}{M_\Sigma}\right) \right],
\end{equation}
where
\begin{equation}
	f(r)=-\frac{r[2r^3 \ln r +(r^2-4)^{\frac{3}{2}}\ln A]}{4},
\end{equation}
\begin{equation}
	A=\frac{(r^2-2-r\sqrt{r^2-4})}{2},
\end{equation}
and $\alpha_2=\frac{g^2}{4\pi}$, with $g=0.65$ and $s_w=0.474$ being the weak mixing angle.
When the mass splitting between the charged and neutral component of the triplet fermion is more than the pion mass, $\Sigma^\pm$ can dominantly decay to $\Sigma^0,\pi^\pm$ and the decay width is given as
\begin{eqnarray}
\Gamma_{\Sigma^{\pm}\rightarrow\Sigma^0\pi^\pm}&=&\frac{2G_F^2}{\pi}f_\pi^2V_{ud}^2(\Delta M)^3\sqrt{1-\frac{m_\pi^2}{(\Delta M)^2}},
\end{eqnarray}
where $G_F=1.166\times10^{-5}~\rm GeV^{-2}$ is the Fermi constant, $f_\pi=131$ MeV is the pion decay constant, $V_{ud}=0.9742\pm0.00021$ is the relevant CKM matrix element, $m_\pi=139.57$ MeV is the pion mass. $\Sigma^\pm$ can also decay through three body leptonic channels $\Sigma^\pm\rightarrow\Sigma^0e\nu_e,\Sigma^\pm\rightarrow\Sigma^0\mu\nu_\mu$.The corresponding decay widths are given by
\begin{eqnarray}   \Gamma_{\Sigma^\pm\rightarrow\Sigma^0e\nu_e}=\frac{2G_F^2(\Delta{M})^5}{15\pi}
\end{eqnarray}
\begin{eqnarray}    \Gamma_{\Sigma^\pm\rightarrow\Sigma^0\mu\nu_\mu}=0.12\Gamma_{\Sigma^\pm\rightarrow\Sigma^0e\nu_e}
\end{eqnarray}
The mixing between the leptons and heavy triplet fermion states can be expressed as \cite{Das:2020uer}
\begin{eqnarray}
    V_{l\Sigma}=M_DM^{-1},
\end{eqnarray}
where \cite{Casas:2001sr}
\begin{eqnarray}
    M_D=U^*_{\rm PMNS}\sqrt{D_m}\mathcal{R}\sqrt{M},
\end{eqnarray}
with $\mathcal{R}$ being a general orthogonal matrix, $D_m$ is the diagonal light neutrino mass matrix with eigen values $m_1$, $\sqrt{\Delta m_{21}^2+m_1^2}$, $\sqrt{\Delta m_{31}^2+m_1^2}$ ($\sqrt{\Delta m_{23}^2-\Delta m_{21}^2+m_3^2}$, $\sqrt{\Delta m_{23}^2+m_3^2}$, $m_3$) in NH (IH) case. $M$ is the diagonal triplet fermion mass matrix with eigen values $M_{\Sigma_1}$, $M_{\Sigma_2}$, and $M_{\Sigma_3}$.
The neutral component of the triplet fermion can decay to the following two body channel with corresponding decay width given as
\begin{eqnarray}  \Gamma_{\Sigma^0\rightarrow l^\pm W^\mp}&=&\frac{g^2|V_{l\Sigma}|^2}{64\pi}\left( \frac{M_\Sigma^3}{M_W^2} \right)\left( 1- \frac{M_W^2}{M_\Sigma^2}\right)^2\nonumber\\&&\left( 1+2 \frac{M_W^2}{M_\Sigma^2}\right)\label{eq:s0tolw}
\end{eqnarray}
\begin{eqnarray}  \Gamma_{\Sigma^0\rightarrow \nu_l Z}&=&\frac{g^2|V_{l\Sigma}|^2}{128\pi}\left( \frac{M_\Sigma^3}{M_W^2} \right)\left( 1- \frac{M_Z^2}{M_\Sigma^2}\right)^2\nonumber\\&&\left( 1+2 \frac{M_Z^2}{M_\Sigma^2}\right)
\end{eqnarray}
\begin{eqnarray}  \Gamma_{\Sigma^0\rightarrow \nu h}&=&\frac{g^2|V_{l\Sigma}|^2}{128\pi}\left( \frac{M_\Sigma^3}{M_W^2} \right)\left( 1- \frac{M_h^2}{M_\Sigma^2}\right)^2
\end{eqnarray}
The charged component of the triplet fermion can decay to the following two body channel with corresponding decay width given as
\begin{eqnarray}  \Gamma_{\Sigma^\pm\rightarrow \nu_l W^\pm}&=&\frac{g^2|V_{l\Sigma}|^2}{32\pi}\left( \frac{M_\Sigma^3}{M_W^2} \right)\left( 1- \frac{M_W^2}{M_\Sigma^2}\right)^2\nonumber\\&&\left( 1+2 \frac{M_W^2}{M_\Sigma^2}\right)
\end{eqnarray}
\begin{eqnarray}  \Gamma_{\Sigma^\pm\rightarrow l^\pm Z}&=&\frac{g^2|V_{l\Sigma}|^2}{64\pi}\left( \frac{M_\Sigma^3}{M_W^2} \right)\left( 1- \frac{M_Z^2}{M_\Sigma^2}\right)^2\nonumber\\&&\left( 1+2 \frac{M_Z^2}{M_\Sigma^2}\right)
\end{eqnarray}
\begin{eqnarray}  \Gamma_{\Sigma^\pm\rightarrow l^\pm h}&=&\frac{g^2|V_{l\Sigma}|^2}{64\pi}\left( \frac{M_\Sigma^3}{M_W^2} \right)\left( 1- \frac{M_h^2}{M_\Sigma^2}\right)^2\label{eq:sptolh}
\end{eqnarray}
By choosing $\mathcal{R}$ to be an identity matrix it has been shown that the above decay modes (Eq. \ref{eq:s0tolw}-\ref{eq:sptolh}) of the 1st generation of triplet are sensitive to the lightest neutrino mass ($m_1$ for NH and $m_3$ for IH) while the decay modes of 2nd and 3rd generation of triplets are independent of lightest neutrino mass \cite{Das:2020uer}.}

\section{First-order phase transition}\label{app:fopt}
The finite temperature effective potential is given as
\begin{eqnarray}
V_{\rm eff}(\phi^\prime,T)&=&V_{\rm tree}(\phi^\prime)+V_{\rm CW}(\phi^\prime)+V_{\rm ct}(\phi^\prime)\nonumber\\&&+V_{T}(\phi^\prime,T)+V_{\rm daisy}(\phi^\prime,T)
\end{eqnarray}
The tree-level part of the potential is given as
\begin{eqnarray}
 V_{\rm tree}=-\frac{1}{2}\lambda_{\phi^\prime}v^2\phi^{\prime^2}+\frac{1}{4}\lambda_{\phi^\prime}\phi{^\prime}^4,
\end{eqnarray}
where $v=\sqrt{3u^2+v_\phi^2}$.
The one-loop Coleman-Weinberg zero temperature contribution in $\rm \overline{MS}$ renormalization scheme is given as
\begin{eqnarray}
V_{\rm CW}=\sum_{i}\frac{n_i}{64\pi^2}m^4_i(\phi^\prime)\bigg(\log\bigg(\frac{m_i^2(\phi^\prime)}{\mu_R^2}\bigg)-c_i\bigg),
\end{eqnarray}
where $m_i(\phi^\prime)$ being the field dependent masses given as
\begin{eqnarray}
m_{\phi^\prime}(\phi^\prime)&=&\sqrt{-\lambda_{\phi^\prime}v^2+3\lambda_{\phi^\prime}\phi^{\prime2}}~;~~ n_{\phi^\prime}=1,\nonumber\\
m_{\eta^\prime}(\phi^\prime)&=&\sqrt{-\lambda_{\phi^\prime}v^2+\lambda_{\phi^\prime}\phi^{\prime2}}~;~~ n_{\eta^\prime}=1,\nonumber\\
M_{Z_{\rm BL}}(\phi^\prime)&=&4.58267g_{\rm BL}\phi^\prime~;~~ n_{Z_{\rm BL}}=3,
\end{eqnarray}
$n_i$ denotes the corresponding d.o.f, and $c_i=3/2$ for scalars and fermions and 5/3 for gauge bosons. The renormalization scale is fixed at $\mu_R\equiv v$. The counter term is
\begin{eqnarray}
V_{\rm ct}=-\frac{\delta\mu^2}{2}\phi{^\prime}^2+\frac{\delta\lambda_{\phi^\prime}}{4}\phi{^\prime}^4,
\end{eqnarray}
which is derived by solving
\begin{eqnarray}
\frac{\partial(V_{\rm CW}+V_{\rm ct})}{\partial\phi^\prime}\bigg|_{\phi^\prime=v}=0,~\frac{\partial^2(V_{\rm CW}+V_{\rm ct})}{\partial\phi{^\prime}^2}\bigg|_{\phi^\prime=v}=0.\nonumber\\
\end{eqnarray}
The finite temperature contribution is given as
\begin{eqnarray}
V_{T}=\frac{T^4}{2\pi^2}\bigg[\sum_{i\in B}n_i J_{B}\bigg(\frac{m_i(\phi^\prime)}{T}\bigg)- \sum_{j\in F}n_j J_{F}\bigg(\frac{m_j(\phi^\prime)}{T}\bigg) \bigg],\nonumber\\
\end{eqnarray}
where
\begin{eqnarray}
J_{B/F}(y)=\int_0^{\infty}x^2\log\big(1\mp e^{-\sqrt{x^2+y^2}}\big)
\end{eqnarray}
The daisy contribution is 
\begin{eqnarray}
V_{\rm daisy}=\frac{T}{12\pi}\sum_kn_k\bigg([m^2_k(\phi^\prime)]^{3/2}-[m^2_k(\phi^\prime)+\Pi_k(T)]^{3/2}  \bigg),\nonumber\\
\end{eqnarray}
where $n_k$ denotes all d.o.f. for scalars and only longitudinal d.o.f. for vector bosons and $\Pi_k's$ are given as
\begin{eqnarray}
\Pi_{\phi^\prime}(T)&=&\Pi_{\eta^\prime}(T)=\bigg(\frac{g_{\rm BL}^2}{2}+\frac{\lambda_{\phi^\prime}}{3}\bigg)T^2\nonumber\\
\Pi_{Z_{\rm BL}}(T)&=&13.2781g_{\rm BL}^2T^2.
\end{eqnarray}

\begin{figure*}[tbh]
\centering
	\begin{tabular}{cc}
    \includegraphics[scale=0.34]{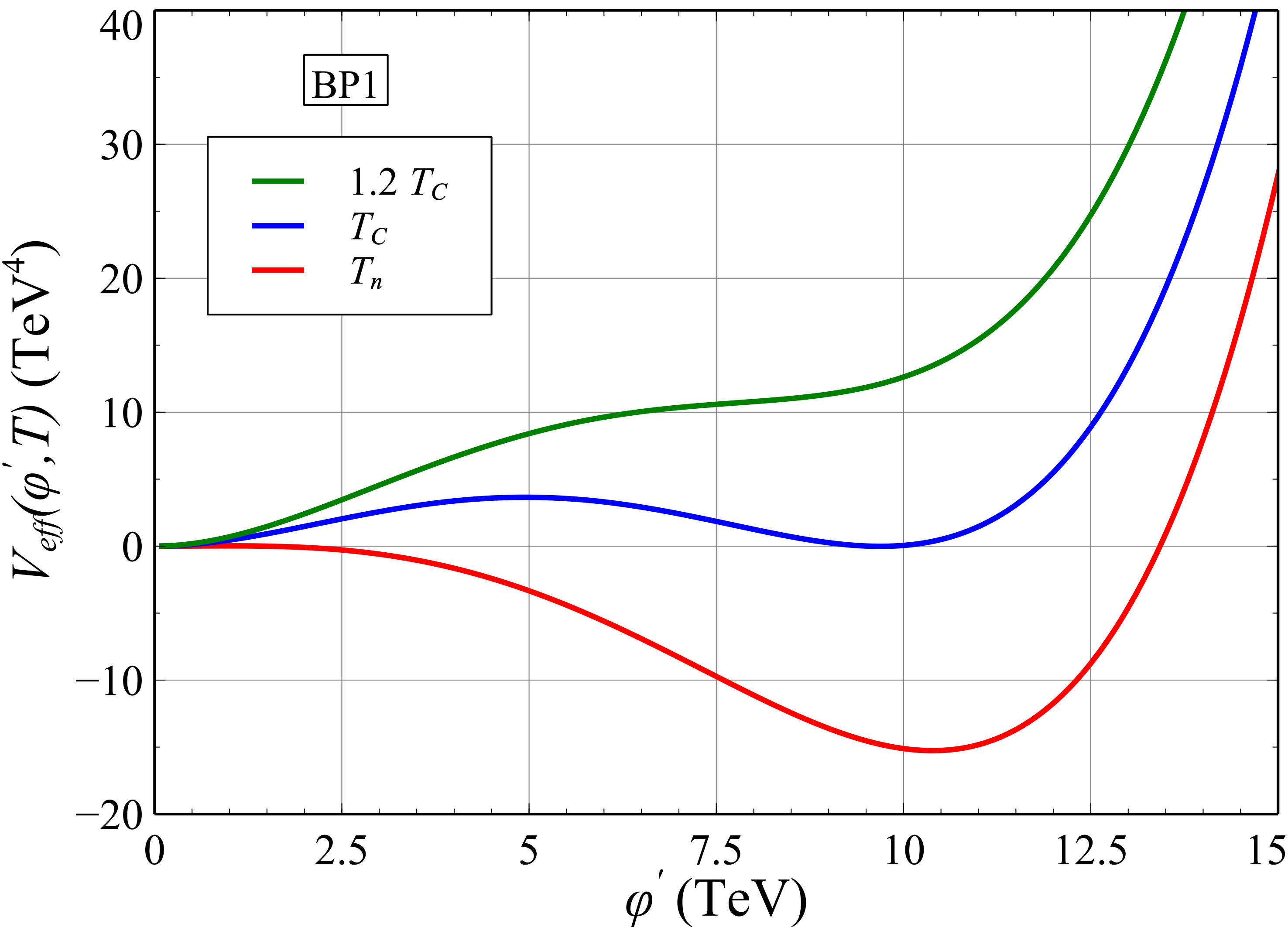}
    \includegraphics[scale=0.34]{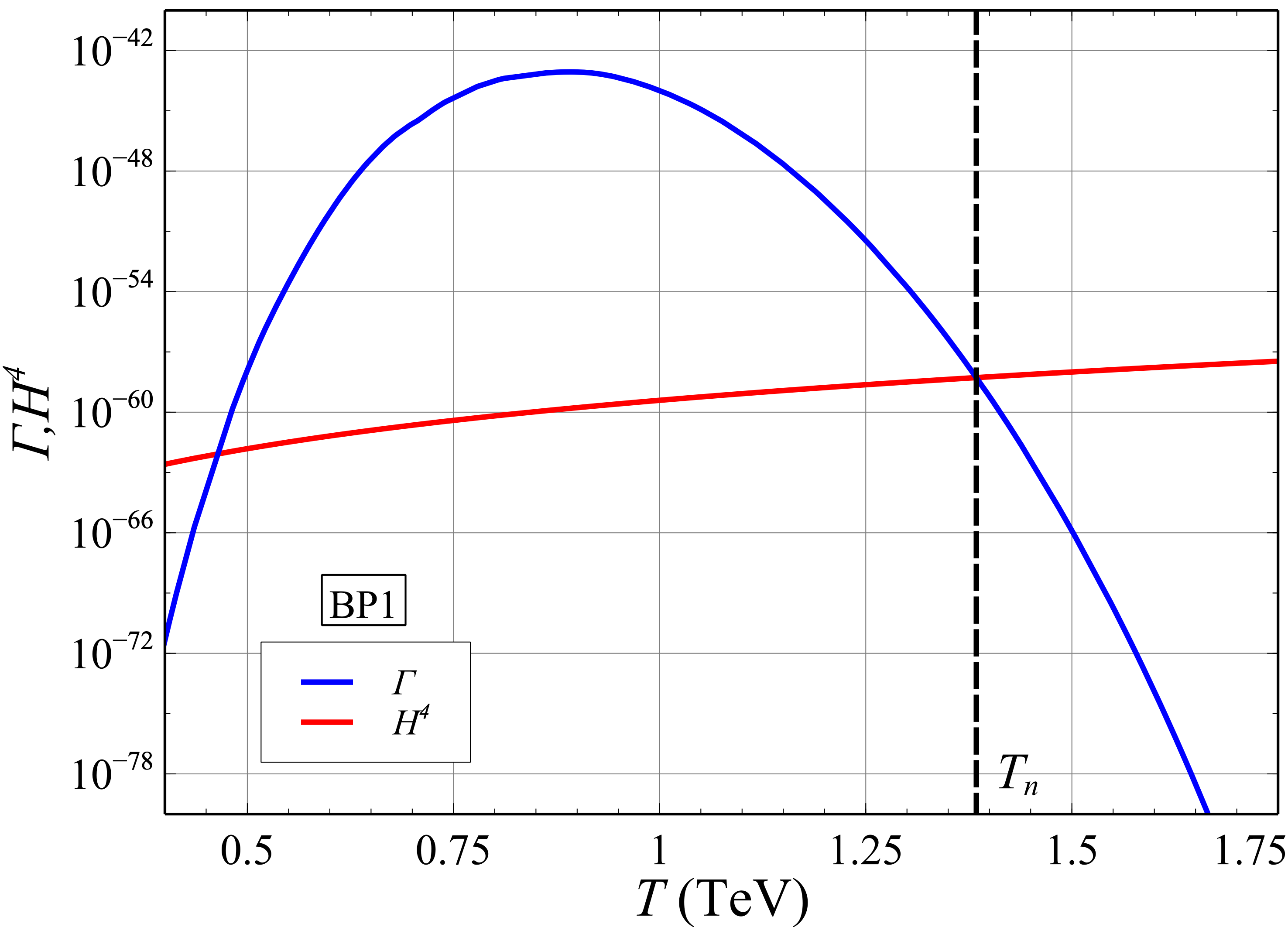}
    \end{tabular}
	\caption{\textit{left}: Effective potential profile for BP1 from Table \ref{tab:tab2} at different temperatures. \textit{right}: Variation of $\Gamma$, and $\mathcal{H}^4$ with $T$ for BP1 from Table \ref{tab:tab2}.}\label{fig:pot_bp1}
\end{figure*}

The bubble nucleation rate per unit volume at a finite temperature is given by
\begin{eqnarray}
\Gamma(T)\simeq \bigg(\frac{S_3(T)}{2\pi T}\bigg)^{3/2}T^4 e^{-\frac{S_3(T)}{T}},
\end{eqnarray}
where $S_3$ is the three-dimensional Euclidean action given as 
\begin{eqnarray}
S_3(T)=\int d^3r\bigg[\frac{1}{2}\big(\frac{d\phi^\prime}{dr}\big)^2+V_{\rm eff}(\phi^\prime,T)\bigg],
\end{eqnarray}
which is calculated by solving
\begin{eqnarray}
\frac{d^2\phi^\prime}{dr^2}+\frac{2}{r}\frac{d\phi^\prime}{dr}-\frac{dV_{\rm eff}(\phi^\prime,T)}{d\phi^\prime}=0,
\end{eqnarray}
with the following boundary conditions
\begin{eqnarray}
\frac{d\phi^\prime}{dr}\bigg|_{r=0}=0, ~\phi^\prime(\infty)=\phi_{\rm false}.
\end{eqnarray}

The nucleation process begins when the bubble nucleation rate becomes comparable to the Hubble expansion rate, which is described by
\begin{eqnarray}
\Gamma(T_n)\approx \mathcal{H}(T_n)^4,
\end{eqnarray}
where $T_n$ is the nucleation temperature, which corresponds to a temperature where $S_3/T$ is approximately 140.

We have illustrated the temperature dependence on the effective potential in the \textit{left} panel of Fig. \ref{fig:pot_bp1} for BP1 from Table \ref{tab:tab2}. The variation of $\Gamma$ and $\mathcal{H}^4$ as a function of temperature is shown in the \textit{right} panel of Fig. \ref{fig:pot_bp1}, which gives the nucleation temperature as $T_n=1.38414$ TeV.

\section{Gravitational waves from FOPT}\label{app:gw}
\textbf{(i) Bubble collision:}

The peak frequency and the amplitude of the GWs generated by the bubble collision are given by
\begin{eqnarray}
f^{{\rm col}}_{\rm peak}&=&1.65\times10^{-5}({\rm Hz})\bigg(\frac{g_{*}(T_n)}{100}\bigg)^{1/6}\bigg(\frac{T_n}{0.1 {~\rm TeV}}\bigg)\nonumber\\&&\frac{0.64}{2\pi}\bigg(\frac{\beta(T_n)}{\mathcal{H}(T_n)}\bigg),
\end{eqnarray}

\begin{eqnarray}
\Omega_{{\rm col}}h^2&=&1.65\times10^{-5}\bigg(\frac{100}{g_{*}(T_n)}\bigg)^{1/3}\bigg(\frac{\mathcal{H}(T_n)}{\beta(T_n)}\bigg)\nonumber\\&&\bigg(\frac{\kappa_\phi(T_n)\alpha(T_n)}{1+\alpha(T_n)}\bigg)^2 \frac{A(a+b)^c}{\bigg(b\big(\frac{f}{f^{{\rm col}}_{\rm peak}}\big)^{-a/c}+a\big(\frac{f}{f^{{\rm col}}_{\rm peak}}\big)^{b/c}\bigg)^c},\nonumber\\
\end{eqnarray}
with efficiency factor 
\begin{eqnarray}
\kappa_\phi=\frac{1}{1+0.715\alpha(T_n)}\bigg(0.715\alpha(T_n)+\frac{4}{27}\sqrt{\frac{3\alpha(T_n)}{2}}\bigg),\nonumber\\
\end{eqnarray}
and $a=1.03,b=1.84,c=1.45,A=5.93\times10^{-2}$ \cite{Lewicki:2022pdb, Athron:2023xlk, Caprini:2024hue}.

\textbf{(ii) Sound waves:}

The peak frequency and the amplitude of the GWs generated by the sound waves are expressed by \cite{Athron:2023xlk} 
\begin{eqnarray}
f^{{\rm sw}}_{\rm peak}&=&8.9\times10^{-6}({\rm Hz})\bigg(\frac{g_{*}(T_n)}{100}\bigg)^{1/6}\bigg(\frac{1}{v_w}\bigg)\bigg(\frac{T_n}{0.1 {~\rm TeV}}\bigg)\nonumber\\&&\bigg(\frac{z_p}{10}\bigg)\bigg(\frac{\beta(T_n)}{\mathcal{H}(T_n)}\bigg)
\end{eqnarray}
where the wall velocity $v_w\sim1$, and $z_p\sim10$.

\begin{eqnarray}
\Omega_{{\rm sw}}h^2&=&2.59\times10^{-6}\bigg(\frac{100}{g_{*}(T_n)}\bigg)^{1/3}\bigg(\frac{\mathcal{H}(T_n)}{\beta(T_n)}\bigg)v_w\Upsilon(T_n)\nonumber\\&& \bigg(\frac{\kappa_{\rm sw}(T_n)\alpha(T_n)}{1+\alpha(T_n)}\bigg)^2 \frac{7^{3.5}\big(\frac{f}{f^{{\rm sw}}_{\rm peak}(T_n)}\big)^3}{\bigg(4+3\big(\frac{f}{f^{{\rm col}}_{\rm peak}(T_n)}\big)^{2}\bigg)^{3.5}}.
\end{eqnarray}
The efficiency factor is \cite{Espinosa:2010hh}
\begin{eqnarray}
\kappa_{{\rm sw}}=\frac{\alpha(T_n)}{0.73+0.083\sqrt{\alpha(T_n)}+\alpha(T_n)}
\end{eqnarray}
The suppression factor is given as
\begin{eqnarray}
\Upsilon=1-\frac{1}{\sqrt{1+2\tau_{{\rm sw}}(T_n)\mathcal{H}(T_n)}},
\end{eqnarray}
where the lifetime of the sound wave is given as \cite{Guo:2020grp}
\begin{eqnarray}
\tau_{{\rm sw}}=\frac{R_{*}(T_n)}{U_f(T_n)},
\end{eqnarray}
where mean bubble separation is given by
\begin{eqnarray}
R_{*}=\frac{(8\pi)^{1/3}v_w}{\beta(T_n)},
\end{eqnarray}
and the rms fluid velocity is given as
\begin{eqnarray}
U_f=\sqrt{\frac{3\kappa_{{\rm sw}}(T_n)\alpha(T_n)}{4(1+\alpha(T_n))}}.
\end{eqnarray}

\textbf{(iii) Turbulence:}

The peak frequency and the amplitude of the GWs generated by the turbulence are expressed by \cite{Caprini:2015zlo, Athron:2023xlk, Caprini:2024hue}
\begin{eqnarray}
f^{{\rm turb}}_{\rm peak}&=&2.7\times10^{-5}({\rm Hz})\bigg(\frac{g_{*}(T_n)}{100}\bigg)^{1/6}\bigg(\frac{1}{v_w}\bigg)\bigg(\frac{T_n}{0.1 {~\rm TeV}}\bigg)\nonumber\\&&\bigg(\frac{\beta(T_n)}{\mathcal{H}(T_n)}\bigg),
\end{eqnarray}

{\scriptsize\begin{eqnarray}
\Omega_{{\rm turb}}h^2&=&3.35\times10^{-4}\bigg(\frac{100}{g_{*}(T_n)}\bigg)^{1/3}\bigg(\frac{\mathcal{H}(T_n)}{\beta(T_n)}\bigg)\bigg(\frac{\kappa_{\rm turb}(T_n)\alpha(T_n)}{1+\alpha(T_n)}\bigg)^2\nonumber\\&& v_w \frac{\big(\frac{f}{f^{{\rm sw}}_{\rm peak}(T_n)}\big)^3}{\bigg(1+\big(\frac{f}{f^{{\rm col}}_{\rm peak}(T_n)}\big)\bigg)^{3.6}\big(1+\frac{8\pi f}{h_*(T_n)}\big)},
\end{eqnarray}}
where the efficiency factor is $\kappa_{{\rm turb}}=0.1\kappa_{{\rm sw}}$ \cite{Caprini:2015zlo}, and the inverse of the Hubble time at the epoch of GW production, red-shifted to today, is expressed as:
\begin{eqnarray}
h_*=1.65\times10^{-5}({\rm Hz})\bigg(\frac{g_{*}(T_n)}{100}\bigg)^{1/6}\bigg(\frac{T_n}{0.1 {~\rm TeV}}\bigg)
\end{eqnarray}

\providecommand{\href}[2]{#2}\begingroup\raggedright\endgroup

\end{document}